\documentclass[aps,amsmath,amssymb,reprint,twocolumn,pra,superscriptaddress,notitlepage,showpacs,tightenlines]{revtex4-1}
\usepackage{exscale}
\usepackage{relsize}
\usepackage{graphicx}
\usepackage{dcolumn}
\usepackage{bm}
\usepackage{textcomp}
\usepackage{siunitx}
\usepackage{appendix}
\usepackage{array}
\usepackage{color}
\usepackage{textcomp}
\usepackage{amsmath}
\usepackage{float}
\usepackage{verbatim}
\usepackage[colorlinks,citecolor=blue,linkcolor=red,hyperindex,CJKbookmarks]{hyperref}

\allowdisplaybreaks[4]

\begin{document}

\title{Manipulation of magnetic systems by quantized surface acoustic wave via piezomagnetic effect}
\author{Yu-Yuan Chen}
\affiliation{School of Integrated Circuits, Tsinghua University, Beijing 100084, China}

\author{Jia-Heng Wang}
\affiliation{School of Integrated Circuits, Tsinghua University, Beijing 100084, China}

\author{Lu Ning Song}
\affiliation{School of Integrated Circuits, Tsinghua University, Beijing 100084, China}

\author{Yu-xi Liu}
\email{yuxiliu@mail.tsinghua.edu.cn}
\affiliation{School of Integrated Circuits, Tsinghua University, Beijing 100084, China}

\date{\today}

\begin{abstract}
The quantized surface acoustic wave (SAW) in the piezoelectric medium has recently been studied, and is used to control electric dipoles of quantum systems via the electric field produced through piezoelectric effect. However, it is not easy and convenient to manipulate magnetic moments directly by the electric field. We here study a quantum theory of SAW in the piezomagnetic medium. We show that the intrinsic properties of the piezomagnetic medium enable the SAW in the piezomagnetic medium to directly interact with magnetic moments of quantum systems via magnetic field induced by piezomagnetic effect. By taking the strip SAW waveguide made of piezomagnetic medium as an example, we further study the coupling strengths between different magnetic quantum systems with magnetic moments and the quantized single-mode SAW in the waveguide. Based on this, we discuss the interaction between magnetic quantum systems mediated by the quantized multi-mode SAW in piezomagnetic waveguide. Our study provides a convenient way to directly control magnetic quantum systems by quantized SAW, and offers potential applications to on-chip information processing based on solid-state quantum systems via quantized acoustic wave.
\end{abstract}
\maketitle

\section{Introduction}
Surface acoustic wave (SAW), a type of mechanical waves, propagates along the surface of elastic medium with wave vectors orthogonal to the normal direction to the surface and is confined near the surface with a depth of about one wavelength~\cite{SAW-Background-Book1,SAW-Background-Book2,SAW-Background-Book3}. Compared with the electromagnetic waves, SAW has about five orders of reduction in the propagation velocity (typically $3\times 10^3$\,m/s for SAW instead of $3\times 10^8$\,m/s for electromagnetic waves). That means, the wavelength of SAW is about five orders smaller than that of electromagnetic waves at the same frequency. These features make SAW have wide range of applications, e.g., modern electronic communication (such as radar, resonator, bandpass filter, and delay lines)~\cite{SAW-Background-Book1,SAW-Background-Book2,SAW-Background-Book3}, microfluid manipulation~\cite{SAW-Fluid-Control1,SAW-Fluid-Control2,SAW-Fluid-Control3}, cell manipulation~\cite{SAW-Cell-Control1,SAW-Cell-Control2,SAW-Cell-Control3}, and sensor~\cite{SAW-Sensor1,SAW-Sensor2,SAW-Sensor3}.

Quantum acoustics~\cite{QuantumAcoustics-Article1-110-PESAW-Quantization,QuantumAcoustics-Article2,QuantumAcoustics-Article3,QuantumAcoustics-Article4}, which is an interdisciplinary research field of quantum mechanics and mechanical wave including SAW, mainly studies the quantum effects of mechanical vibrations and the interaction between quantized mechanical vibrations and various quantum systems including superconducting circuits~\cite{SAW-Superconducting-Coupling1,SAW-Superconducting-Coupling2,SAW-Superconducting-Coupling3}, trapped ions~\cite{AW-Ions-Coupling1,AW-Ions-Coupling2,SAW-Ions-Coupling1}, defect centers in diamond~\cite{NVcenter-PESAW-Device,SAW-NVcenter-Coupling1,SAW-NVcenter-Coupling2}, and quantum dots~\cite{SAW-QuantumDot-Coupling1,SAW-QuantumDot-Coupling2,SAW-QuantumDot-Coupling3}. A phonon is a quantum of vibrational mechanical energy in the quantized mechanical vibrations. We note that superconducting circuits~\cite{Superconducting-Circuit1,Superconducting-Circuit2,Superconducting-Circuit3,Superconducting-Circuit4} have become one of the promising scalable solid-state platforms for realizing quantum computers. With the rapid development of fabrication for SAW devices and superconducting circuits, attentions are paid to achieve the strong coupling between superconducting circuits and quantized single-mode SAW at single-phonon level. Over the past few years, the quantized single-mode SAW has been experimentally and theoretically studied for quantum entanglement~\cite{SAW-Entanglement1,SAW-Entanglement2,SAW-Entanglement3}, quantum transduction~\cite{SAW-Transducer1,SAW-Transducer2,SAW-Transducer3}, quantum routing~\cite{SAW-Quantum-Routing1,SAW-Quantum-Routing2,SAW-Quantum-Routing3}, phononic blockade~\cite{SAW-Phonon-Blockade1,SAW-Phonon-Blockade2,SAW-Phonon-Blockade3}, and interaction with giant atoms~\cite{SAW-Giant-Acoustic-Atom1,SAW-Giant-Acoustic-Atom2,SAW-Giant-Acoustic-Atom3}.

So far, most existing researches in quantum acoustics focus on the quantization of SAW in the piezoelectric medium~\cite{QuantumAcoustics-Article1-110-PESAW-Quantization}, as well as the coupling between quantum systems and a single-mode (or multi-mode) SAW in the resonator, which is formed by two reflecting gratings fabricated on the surface of the piezoelectric medium~\cite{QuantumAcoustics-Article1-110-PESAW-Quantization,QuantumAcoustics-Article2,QuantumAcoustics-Article3,QuantumAcoustics-Article4,SAW-Superconducting-Coupling1,SAW-Superconducting-Coupling2,SAW-Superconducting-Coupling3,NVcenter-PESAW-Device,SAW-NVcenter-Coupling1,SAW-NVcenter-Coupling2,SAW-QuantumDot-Coupling1,SAW-QuantumDot-Coupling2,SAW-QuantumDot-Coupling3,SAW-Ions-Coupling1,SAW-Entanglement1,SAW-Entanglement2,SAW-Entanglement3,SAW-Transducer1,SAW-Transducer2,SAW-Transducer3,SAW-Quantum-Routing1,SAW-Quantum-Routing2,SAW-Quantum-Routing3,SAW-Phonon-Blockade1,SAW-Phonon-Blockade2,SAW-Phonon-Blockade3,SAW-Giant-Acoustic-Atom1,SAW-Giant-Acoustic-Atom2,SAW-Giant-Acoustic-Atom3}. Thus, the coupling of SAW to quantum systems is realized via the electric field produced through the piezoelectric effect. However, the magnetic moments of magnetic quantum systems (e.g., superconducting qubit, ferromagnetic magnon, and defect center in diamond) cannot be directly coupled to SAW via piezoelectric effect. To solve this problem, one possible solution is to deposite a magnetic film on the surface of the piezoelectric medium~\cite{SAW-Magnetic-Film1,SAW-Magnetic-Film2,SAW-Magnetic-Film3,SAW-Magnetic-Film4}. Then, SAW can be coupled to magnetic quantum systems via magnetic field produced by the magnetic film through magnetoelastic effect. However, this film would inevitably lead to additional decay of SAW and make it difficult to achieve strong coupling.

In contrast to the SAW in piezoelectric medium, the SAW in piezomagnetic medium~\cite{PM-Eq-parameters,PM-Eq-Book1,PM-Eq-Book2-Quasi-Statics} could offer a direct way to couple SAW with magnetic quantum systems via the magnetic field induced by piezomagnetic effect. The quantum theory of SAW in the piezoelectric medium has been developed~\cite{QuantumAcoustics-Article1-110-PESAW-Quantization}, however, the detailed study about the quantum theory of SAW in the piezomagnetic medium is not found. Thus, we here use similar way as in Ref.~\cite{QuantumAcoustics-Article1-110-PESAW-Quantization} and employ the canonical quantization method~\cite{Canonical-Quantization1,Canonical-Quantization2} to study the quantum theory of SAW in the piezomagnetic medium. The derivation is based on the classical SAW in piezomagnetic medium, which is obtained by solving the general wave equations in strip waveguide made of piezomagnetic medium. Based on this, we study the interaction between several magnetic quantum systems and the quantized single-mode SAW in piezomagnetic strip waveguide. Furthermore, we study the interaction between two qubits mediated by quantized multi-mode SAW in piezomagnetic strip waveguide.

The paper is organized as follows. In Sec.~\ref{Sec2-SAW-Classics}, for completeness of the paper, we first give a brief summary for the general acoustic wave equations and boundary conditions in piezomagnetic medium, and derive the typical Rayleigh-type SAW in an piezomagnetic strip waveguide. In Sec.~\ref{Sec3-SAW-Quantum}, we resort to the canonical quantization method to derive the quantum theory of SAW in piezomagnetic strip waveguide and discuss the zero-point fluctuation of the quantized SAW. In Sec.~\ref{Sec4-SAW-Coupling-to-Systems}, the interaction between several typical magnetic quantum systems and a quantized single-mode SAW in the strip waveguide is studied, the coupling strengths of the magnetic quantum systems to a quantized single-mode SAW at single-phonon level are comparatively summarized. As further applications, in Sec.~\ref{Sec5-SAW-mediated-Interaction}, we study the information transfer and entanglement between two superconducting qubits mediated by the SAW phonon in the piezomagnetic waveguide. Finally, the conclusion is given in Sec.~\ref{Sec6-Conclusion}. For conciseness and completeness of the paper, detailed derivations are provided in the Supplementary Material.

\section{Classical SAW in piezomagnetic medium}
\label{Sec2-SAW-Classics}

\begin{figure}[tb]
	\centering
	\includegraphics[scale=0.16]{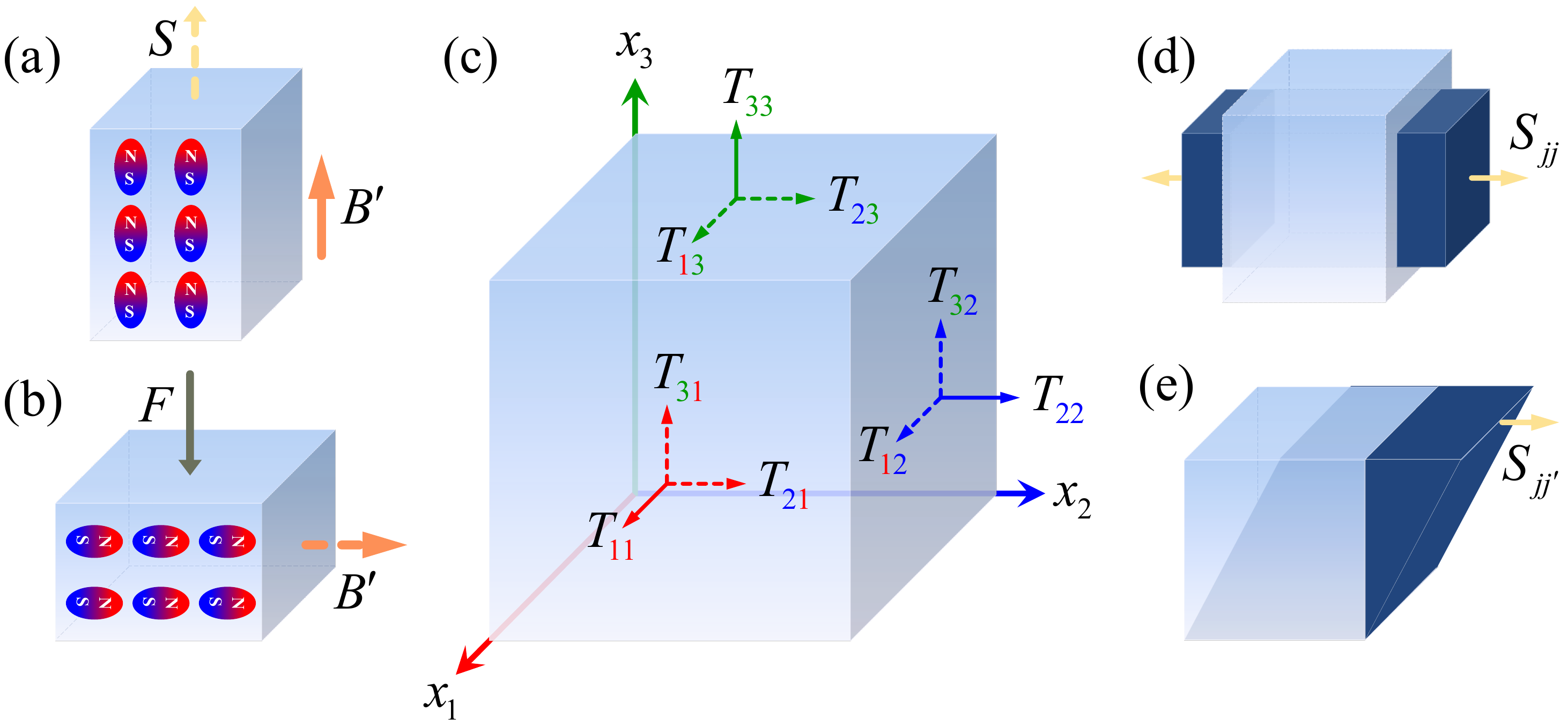}
	\caption{(a) Schematic illustration for the (a) converse piezomagnetic effect and (b) piezomagnetic effect. (a) The externally applied magnetic field $B'$ results in the mechanical strain $S$ and the align of magnetic moments inside the medium. (b) The mechanical strain $S$ and the align of magnetic moments, produced by the externally applied force $F$, leads to the generation of magnetic field $B'$. The relation between the magnetic field $B$ inside the medium and the magnetic field $B'$ outside the medium is determined by the magnetic boundary condition in Eq.~(\ref{PotentialBoundary}). (c) Schematic illustration for the normal stresses (solid arrows) and shear stresses (dashed arrows) acting on a unit cube. (c) Schematic illustration for the deformation of the (d) normal strains $S_{jj}$ and (e) shear strains $S_{jj'}$~($j'\neq j$) of a unit cube.}
	\label{F1-PiezomagneticEffect}
\end{figure}

Within the piezoelectric medium, the mechanical displacements or strains would induce electric fields via piezoelectric effect. In a similar manner, within the piezomagnetic medium, the mechanical displacements or strains would lead to the generation of magnetic fields via piezomagnetic effect as schematically shown in Fig.~\ref{F1-PiezomagneticEffect}(a). In this section, for the completeness of the studies, we summarize several main results for classical SAW in the strip waveguide made of the piezomagnetic medium.

\subsection{Acoustic wave equations in piezomagnetic medium and boundary conditions}

In the following, we limit our discussions to the case of linear piezomagnetic coupling, which determines the basic properties of SAW. In the piezomagnetic medium, the propagation of acoustic waves is described by the coupled constitutive equations that connect the mechannical stress and magnetic induction as follows~\cite{PM-Eq-parameters,PM-Eq-Book1}
\begin{align}
&T_{ij}=C_{ijkl}S_{kl} -q_{kij}H_k,\nonumber\\
&B_i=\mu_{ik}H_k +q_{ijk}S_{jk},
\label{PiezomagneticEq}
\end{align}
where the first subscript of the stress tensor $T_{ij}$ denotes the direction of the force exerting the area under consideration, and the second one denotes the outward going normal direction of the area that the force is exerted, as schematically shown in Fig.~\ref{F1-PiezomagneticEffect}(b). Each subscript can take $1$, $2$, or $3$, which denote $x$, $y$, or $z$ direction, respectively. $B_i$ is the magnetic induction along $x_i$ direction. $C_{ijkl}$, $q_{kij}$, and $\mu_{ik}$ represent the elastic stiffness, piezomagnetic, and magnetic permeability tensors, respectively. $S_{jk}=\left({\partial u_{j}}/{\partial x_{k}} +{\partial u_{k}}/{\partial x_{j}} \right) /2$ is a component of the strain $S$ determined by the mechanical displacements $u_{j}$ and $u_{k}$ that are, respectively, along $x_j$ and $x_k$ directions as schematically shown in Fig.~\ref{F1-PiezomagneticEffect}(c). $H_k$ denotes the magnetic field strength $H$ along $x_k$ direction. Here, the magnetic field $H_k=-{\partial \psi}/{\partial x_{k}}$ is determined by only the magnetic potential $\psi$ under the quasi-static approximation~\cite{PM-Eq-Book2-Quasi-Statics,PM-Eq-Book3-Quasi-Statics}, wherein the piezomagnetic coupling between acoustic and electromagnetic waves is negligible in comparison with the effect of piezomagnetic stiffening.

According to the coupled constitutive equation in Eq.~(\ref{PiezomagneticEq}) and the motion equation $\rho\,{\partial^2 u_i} \big/ {\partial t^2} ={\partial T_{i j}} \big/ {\partial x_j}$ of particle displacement in a solid without body force~\cite{SAW-Background-Book1,SAW-Background-Book3,PM-Eq-Book1}, one can obtain
\begin{equation}
\rho \frac{\partial^2 u_i}{\partial t^2}=
C_{ijkl}\frac{\partial S_{kl}}{\partial x_{j}}
-q_{kij}\frac{\partial H_k}{\partial x_{j}},
\label{DisplaceEQ}
\end{equation}
where $\rho$ denotes the mass density of the medium. For a piezomagnetic medium, the magnetic induction $B$ satisfies
\begin{equation}
\sum_{i=1}^{3}
\left(
\mu_{ik}\frac{\partial H_k}{\partial x_{i}}
+q_{ijk}\frac{\partial S_{jk}}{\partial x_{i}}
\right)
=0,
\label{PotentialEQ}
\end{equation}
which is derived from Eq.~(\ref{PiezomagneticEq}) by using $\nabla \cdot B = 0$~\cite{EM-Book}.

Here, we consider that the surface $z=0$ of the piezomagnetic medium is stress free. That means, the three components of stress exerted on the surface $z=0$ should vanish, i.e.,
\begin{equation}
T_{13}=T_{23}=T_{33}=0.
\label{MechanicalBoundary}
\end{equation}
The magnetic induction $B$ must be continuous across the surface of the medium~\cite{EM-Book}, i.e., $B_3(z=0^+)=B_3(z=0^-)$. By taking the exponentially decaying magnetic potential~\cite{QuantumAcoustics-Article1-110-PESAW-Quantization,EM-Book} ${\psi}'= {\psi}'_{0} e^{kz}$ and magnetic field induction $B'=-\mu_0 {\partial {\psi}'}/{\partial z}$ in the vacuum half-space outside the medium (we here assume that this space is in the direction $z<0$), one can obtain the magnetic boundary condition given as
\begin{equation}
\mu_0 k \psi +\mu_{3k}H_k +q_{3jk}S_{jk}=0.
\label{PotentialBoundary}
\end{equation}

\subsection{Rayleigh-type SAW in isotropic piezomagnetic strip waveguide}

The Rayleigh-type SAW, which stands as a typical SAW~\cite{QuantumAcoustics-Article1-110-PESAW-Quantization,Rayleigh-SAW-Original,Rayleigh-SAW-Review,Rayleigh-SAW-Book}, contains a longitudinal vibration and a transverse vibration, with a phase difference $\pi/2$ between them. Hereafter, by taking the strip waveguide made of piezomagnetic medium as an example, we consider the travelling-wave Rayleigh-type SAW propagating in [110] direction for convenience. Thus, the ansatz of the mechanical displacement is given as~(see Sec.~I of the Supplemental Material in detail)
\begin{align}
u_k=\sum_{k}
	\bigg(\begin{array}{c}
		u'_{1,k} \\
		u_{3,k}
	\end{array}\bigg)=
	\sum_{k}
	\bigg(\begin{array}{c}
		U'_k \\
		iW_k
	\end{array}\bigg) e^{-kq z} e^{i(k x' -\omega_k t)},
\label{110-Ansatz}
\end{align}
where $k=2\pi\big/\lambda_k$ denotes the wavenumber of SAW with the wavelength $\lambda_k$. $\sum_{k}$ denotes the summation over all allowed values of $k$ range from negative infinity to positive infinity. $u'_{1,k}$ and $u_{3,k}$ denotes the components of the mechanical displacement, with the amplitudes $U'_k$ and $W_k$. $q$ represents the exponential decay of SAW into the bulk. $\omega_k=k v$ denotes the frequency of SAW, with the propagation velocity $v$. Here, $u'_{1,k}/\sqrt{2}=u_{1,k}=u_{2,k}$ is the displacement component along the defined direction $x'=(x_1+x_2)/\sqrt{2}$, with $x'$ corresponding to the [110] direction of crystal.

We here take an isotropic peizomagnetic sample $\text{terfenol}$-$\text{D}$~\cite{Parameter-terfenolD}, which is a widely used peizomagnetic material, as an example for the following discussions. For this sample, the corresponding elastic stiffness, piezomagnetic and magnetic permeability tensors have only the independent nonzero constants $C_{11}$, $C_{12}$, $C_{44}$, $q_{31}$, $q_{33}$, and $\mu_{11}$. Thus, by including the normalization condition for the ansatz in Eq.~(\ref{110-Ansatz}) and the detailed structure of the peizomagnetic medium, we obtain the solution of mechanical displacement in the waveguide as follows~(see Sec.~II of the Supplemental Material in detail)
\begin{align}
u=&\sum_{k} 2U_{0,k}
    \bigg(\begin{array}{c}
    \text{cos}\left(k q_{\beta} z +\theta\right)\\
    -i\left|\gamma\right|\text{cos}\left(q_{\beta} k z +\theta +\xi\right)
    \end{array}\bigg)
    e^{-k q_{\alpha}z}
    e^{i(k x' -\omega_k t)}.
\label{uk-solution}
\end{align}
with the normalization constant $U_{0,k}$, which will be further discussed in the next section.

In terms of the mechanical displacement in Eq.~(\ref{uk-solution}), by solving the coupled equation in Eq.~(\ref{PotentialEQ}) and considering magnetic boundary condition in Eq.~(\ref{PotentialBoundary}), we obtain the solution of magnetic potential in the waveguide given as~(see Sec.~II of the Supplemental Material in detail)
\begin{align}
\psi=
\begin{cases}
		\sum_{k} i\psi_{0,k} F_k\left(z\right) e^{i(kx' -\omega_k t)}, &z \geq 0 \\
		\sum_{k} i\psi_{0,k} F_k\left(z=0\right) e^{kz} e^{i(kx' -\omega_k t)}, &z<0
\end{cases},
\label{Psi-solution}
\end{align}
where $\psi_{0,k}=q_{33} U_{0,k} \big/ \mu_{11}$ is the amplitude of the magnetic potential. $F_k\left(z\right)=2A e^{-k q_{\alpha}z} \text{cos}\left(k q_{\beta}z+\theta+\tau\right) +A_3 e^{-kz}$ describes the depth scale on which the magnetic potential of SAW decays into the bulk.

\begin{table}
\centering
\caption{Elastic, piezomagnetic, and magnetic permeability properties for material~$\text{terfenol}$-$\text{D}$~\cite{Parameter-terfenolD}.}
\renewcommand{\arraystretch}{1.8}
\begin{tabular}
	{p{1.8cm}<{\centering}|p{1.8cm}<{\centering}|p{1.8cm}<{\centering}|p{1.8cm}<{\centering}}
	\hline
	$\rho\,(\text{g/cm}^3)$  &$C_{11}$\,(GPa)  &$C_{12}$\,(GPa) &$C_{44}$\,(GPa) \\
	\hline		
	$9.06$  &$55$  &$43$  &$12$  \\
	\hline
	$q_{31}$\,($\text{N/Am}$)  &$q_{33}$\,($\text{N/Am}$) &$\mu_{11}$\,($\SI{}{\micro\newton} /$A$^{2}$) \\
	\hline		
	$-45$  &$90$  &$6.283$ \\
	\hline
\end{tabular}
\label{Material-Parameters}
\end{table}

\begin{figure}[tbp]
\centering
	\includegraphics[width=8.5cm]{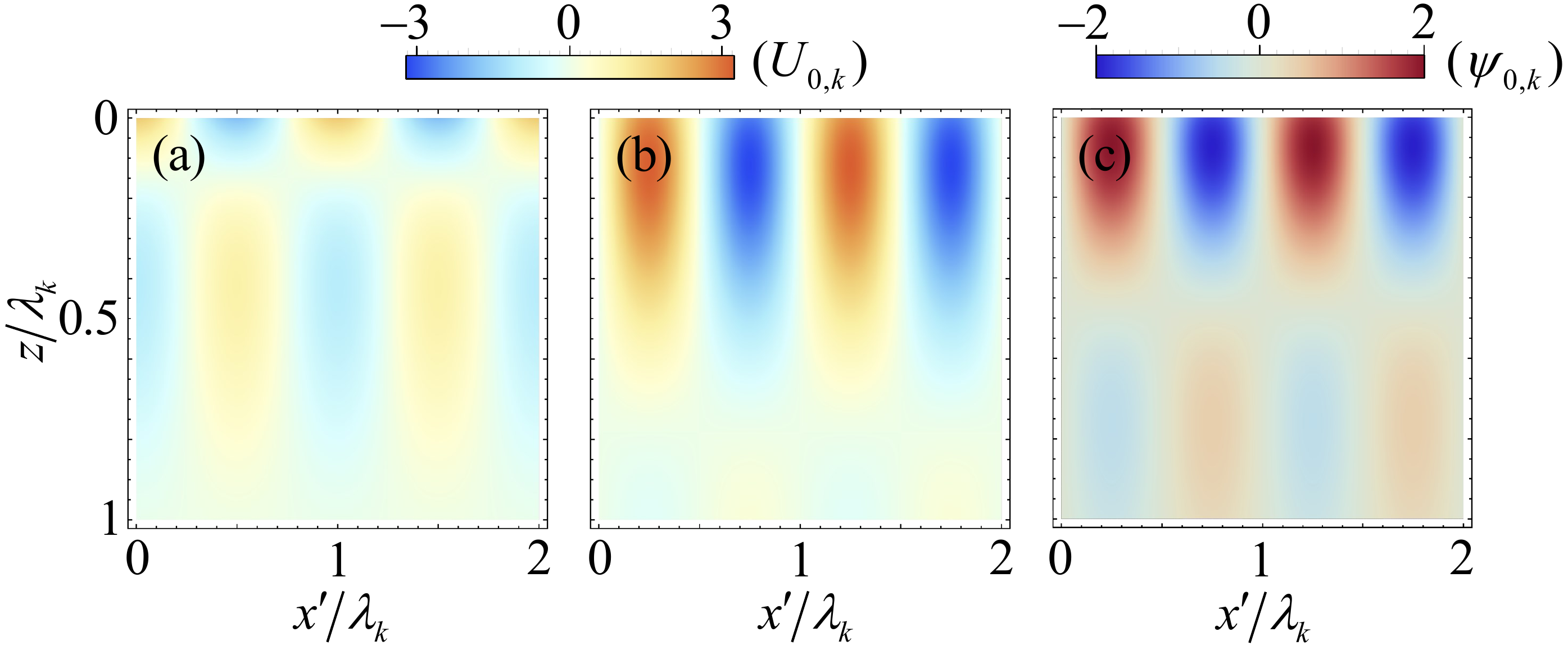}
	\caption{The spatial distribution of the (a) longitudinal mechanical displacement $u'_{1,k}$, (b) transverse mechanical displacement $u_{3,k}$, and (c) magnetic potential $\psi_k$ for the SAW with wavelength $\lambda_k$. Other parameters are $\rho=9.06\,\text{g}/\text{cm}^3$, $C_{11}=55\,\text{GPa}$, $C_{12}=43\,\text{GPa}$, $C_{44}=12\,\text{GPa}$, $q_{31}=-45\,\text{N}/\text{Am}$, $q_{33}=90\,\text{N}/\text{Am}$, and $\mu_{11}=6.283\,\SI{}{\micro\newton}/\text{A}^2$.}
\label{F2-SAW-xz-dependence}
\end{figure}

From the parameters of isotropic $\text{terfenol}$-$\text{D}$~\cite{Parameter-terfenolD} given in Table~\ref{Material-Parameters}, the parameters in the solutions of mechanical displacement in Eq.~(\ref{uk-solution}) and magnetic potential in Eq.~(\ref{Psi-solution}) are calculated as
\begin{align}
&v=1005\,\text{m/s},~q_{\alpha}=0.4288,~q_{\beta}=0.5378,
	\nonumber\\
&\theta=1.0700,~\left|\gamma\right|=1.4116,~\xi=-2.1401,
	\nonumber\\
&A=0.8437,~\tau=1.9172,~A_3=1.0370.
\end{align}
In order to study the properties of the Rayleigh-type SAW in piezomagnetic waveguide, the dependence of mechanical displacement and magnetic potential on the spatial positions $x'$ and $z$ is shown in Fig.~\ref{F2-SAW-xz-dependence}. It is found that both the mechanical displacement and magnetic potential decay rapidly along $+z$ direction, and their corresponding amplitudes approach zero near the depth of one wavelength. That is, the energy of SAW is confined to near the surface of the medium. For the mechanical displacement of SAW, there exists a phase difference $\pi/2$ between the longitudinal and transverse components as shown in Figs.~\ref{F2-SAW-xz-dependence}(a)~and~\ref{F2-SAW-xz-dependence}(b). Thus, for any particle within the medium, as the SAW propagates, it undergoes elliptical motion in the vicinity of its equilibrium position. Additionally, there is a phase difference $\pi /2$ between the longitudinal mechanical displacement and magnetic potential as shown in Figs.~\ref{F2-SAW-xz-dependence}(a)~and~\ref{F2-SAW-xz-dependence}(c). Note that the properties exhibited here are similar to those of the Rayleigh-type SAW in piezoelectric medium~\cite{SAW-Background-Book1,SAW-Background-Book2,SAW-Background-Book3,QuantumAcoustics-Article1-110-PESAW-Quantization}.

\section{Quantization of SAW In Piezomagnetic Waveguide}
\label{Sec3-SAW-Quantum}

We have obtained the solutions of mechanical displacement and magnetic potential for classical Rayleigh-type SAW in piezomagnetic waveguide. Let us now apply the canonical quantization method~\cite{Canonical-Quantization1,Canonical-Quantization2} to achieve the quantization of SAW in piezomagnetic waveguide. To show the quantization process clearly, we take the Rayleigh-type SAW obtained in the last section as an example. Here, we point out that the employed canonical quantization method is also applicable to the quantization of other type of SAWs in other piezomagnetic materials.

\subsection{Lagrangian of SAW}
The Lagrangian of SAW propagating in the piezomagnetic medium is written as
\begin{align}
\mathcal{L}&=E_{\bm{k}} -E_{\bm{p}},
\label{Lagrangian}
\end{align}
where $E_{\bm{k}}$ and $E_{\bm{p}}$ denote, respectively, the kinetic energy and potential energy terms given as
\begin{align}
E_{\bm{k}}&=\int \frac{1}{2} \rho
    \left(\frac{\partial u}{\partial t}\right)^2 dV,
    \\
E_{\bm{p}}&=\int \frac{1}{2} \left(S\cdot T -H\cdot B\right) dV.
\end{align}
In terms of the generalized coordinates $u'_1$, $u_3$ and $\psi$, the detail of the kinetic energy and potential energy terms are given as
\begin{align}
E_{\bm{k}}=\,& \mathlarger{\int} \frac{1}{2} \rho
    \left[
	\left(\frac{\partial u'_1}{\partial t}\right)^2+
	\left(\frac{\partial u_3}{\partial t}\right)^2
	\right] dV,
	\nonumber\\
E_{\bm{p}}=\,&\mathlarger{\int} \frac{1}{2}
    \bigg[C'_{11}\bigg(\frac{\sqrt{2}\partial u'_1}{\partial x_1}\bigg)^2+
    C_{44}\bigg(\frac{\partial u'_1}{\partial x_3}+\frac{\sqrt{2}\partial u_3}{\partial x_1}\bigg)^2
    \bigg] dV+
    \nonumber\\
	&\mathlarger{\int} \frac{1}{2}
	\bigg[
	C_{11}\bigg(\frac{\partial u_3}{\partial x_3}\bigg)^2+
	2C_{12} \frac{\sqrt{2}\partial u'_1}{\partial x_1}
	\frac{\partial u_3}{\partial x_3}
	\bigg] dV-
	\nonumber\\
	&\mathlarger{\int} \frac{1}{2}
	\bigg[
	\mu_{11}\bigg(\frac{\sqrt{2}\partial \psi}{\partial x_1}\bigg)^2+
	\mu_{11}\bigg(\frac{\partial \psi}{\partial x_3}\bigg)^2
	\bigg] dV-
	\nonumber\\
	&\mathlarger{\int}
	\bigg(
	q_{31}\frac{\sqrt{2}\partial u'_1}{\partial x_1} \frac{\partial \psi}{\partial x_3}
	+q_{33}\frac{\partial u_3}{\partial x_3} \frac{\partial \psi}{\partial x_3}
	\bigg) dV.
\label{Lagrangian-Detail}
\end{align}

\subsection{Canonical coordinates and momentums of SAW}
For the Rayleigh-type SAW, there exists a phase difference $\pi/2$ between $u'_1$ and $u_3$, as well as $u'_1$ and $\psi$. If $u'_1$, $u_3$ and $\psi$ are chosen as the canonical coordinates, such a phase difference would lead to the non-commutation between different coordinates. Therefore, $u'_1$, $u_3$ and $\psi$ are a set of generalized coordinates rather than canonical coordinates.

In our study, the quantization of SAW is achieved via the canonical quantization method~\cite{Canonical-Quantization1,Canonical-Quantization2}, which uses the canonical coordinates and canonical momentums. For a given system, the canonical coordinates and corresponding canonical momentums are not unique, but they are required to satisfy the equations of motion and the fundamental Poisson bracket relations. Therefore, we consider to transform the generalized coordinates $u'_1$, $u_3$ and $\psi$ to new coordinates such that the new coordinates commutate to each other in the Lagrangian.

The spatial derivatives of the mechanical displacements and magnetic potential are given in terms of themselves as~(see Sec.~III of the Supplemental Material in detail)
\begin{align}
\frac{\partial u'_{1}(\bm{r})}{\partial x_1}=&
    \sum_{k>0} k g_{k}^{(1)} u_{3,k}(\bm{r}),
~\frac{\partial u'_{1}(\bm{r})}{\partial x_3}=
    \sum_{k>0} k g_{k}^{(2)} u'_{1,k}(\bm{r}),
\nonumber\\
\frac{\partial u_{3}(\bm{r})}{\partial x_1}=&
    \sum_{k>0} k g_{k}^{(3)} u'_{1,k}(\bm{r}),
~\frac{\partial u_{3}(\bm{r})}{\partial x_3}=
    \sum_{k>0} k g_k^{(4)} u_{3,k}(\bm{r}),
\nonumber\\
\frac{\partial \psi(\bm{r})}{\partial x_1}=&
    \sum_{k>0} k g_k^{(5)} u'_{1,k}(\bm{r}),
~\frac{\partial \psi(\bm{r})}{\partial x_3}=
    \sum_{k>0} k g_k^{(6)} \psi_k(\bm{r}),
\label{partial-represented-by-coordinate}
\end{align}
where the term $(\bm{r})$ denotes the spatial term of the generalized coordinates. The summation $\sum_{k>0}$ is over allowed values of $k$ range from zero to positive infinity. To find the canonical coordinates transformed from the generalized coordinates $u'_1$, $u_3$ and $\psi$, we substitute Eq.~(\ref{partial-represented-by-coordinate}) into Eq.~(\ref{Lagrangian-Detail}). Thus, the kinetic energy and potential energy terms can be rewritten as
\begin{align}
E_{\bm{k}}=&\mathlarger{\int} \sum_{k>0} \frac{1}{2} \rho
    \left(\frac{\partial u'_{1,k}}{\partial t},\frac{\partial u_{3,k}}{\partial t}\right)
    \left(\frac{\partial u'_{1,k}}{\partial t},\frac{\partial u_{3,k}}{\partial t}\right)^T dV,
\nonumber\\
E_{\bm{p}}=&\mathlarger{\int} \sum_{k>0}
    \frac{1}{2} k^2 \left(u'_{1,k},u_{3,k},\psi_k\right) G_k(z)
    \left(u'_{1,k},u_{3,k},\psi_k\right)^T
    dV,
\label{PotetialEnergy-Detial}
\end{align}
where $G_k(z)$, given in Sec.~III of the Supplemental Material, is the quadratic-form coefficient matrix constructed according to Eq.~(\ref{partial-represented-by-coordinate}).

To eliminate the cross terms of coordinates in the potential energy term in Eq.~(\ref{PotetialEnergy-Detial}), we take the orthogonal matrix $Q_{k}^{(1)}$ to diagonalize $G_{k}(z)$ such that
\begin{align}
\lambda_{k}^{(1)}
    &\equiv{\rm diag} \big(\lambda_{1,k}^{(1)},\,\lambda_{2,k}^{(1)},\,\lambda_{3,k}^{(1)}\big)
    \nonumber\\    
    &=Q_{k}^{(1)T} G_{k}(z) Q_{k}^{(1)},
\end{align}
where $Q_{k}^{(1)}$ is constructed via the eigenvector of the quadratic-form coefficient matrix $G_{k}(z)$. The eigenvalue $\lambda_{j,k}^{(1)}$ is to the root of the secular equation $\det\left(G_{k}(z) -\lambda\bm{\text{I}} \right)=0$, with the identity matrix $\bm{\text{I}}$. Therefore, the potential energy term in Eq.~(\ref{PotetialEnergy-Detial}) is written as
\begin{align}
E_{\bm{p}}=&\mathlarger{\int}\sum_{k>0} \frac{1}{2} k^2
    X_{k}^{(1)T} Q_{k}^{(1)T} G_{k}(z) Q_{k}^{(1)} X_{k}^{(1)} dV,
\end{align}
where $X_{k}^{(1)}= \big[Q_{k}^{(1)}\big]^{-1} \big(u'_{1,k},\,u_{3,k},\,\psi_k\big)^T$ denotes the coordinate transformed from the generalized coordinates $u'_{1,k}$, $u_{3,k}$ and $\psi_{k}$ via the orthogonal matrix $Q_{k}^{(1)}$. However, we note that $X_{k}^{(1)}$ is not the expected canonical coordinate since the transform $Q_{k}^{(1)}$ would lead to the cross terms of coordinates in the kinetic energy term. That is,
\begin{align}
E_{\bm{k}}&=\mathlarger{\int} \sum_{k>0} \frac{1}{2} \rho
    \bigg[\frac{\partial X_{k}^{(1)}}{\partial t}\bigg]^T
    Y_k(z)
    \bigg[\frac{\partial X_{k}^{(1)}}{\partial t}\bigg] dV,
\end{align}
where $Y_k(z)$, given in Sec.~III of the Supplemental Material, is constructed from the matrix $Q_{k}^{(1)}$. Hence, it is necessary to further consider the coordinate transform via the orthogonal matrix $Q_k^{(2)}$, which ensures that no cross terms of transformed coordinates exists in both the kinetic energy and potential energy terms. Thus, we can construct the orthogonal matrix $Q_k^{(2)}$ via the normalized eigenvector of $\big[\lambda_{k}^{(1)}\big]^{-1} Y_k$, that is,
\begin{align}
\lambda_{k}^{(2)}
    &\equiv{\rm diag} \big(\lambda_{1,k}^{(2)},\,\lambda_{2,k}^{(2)},\,\lambda_{3,k}^{(2)}\big)
    \nonumber\\
    &=Q_k^{(2)T} \big[\lambda_{k}^{(1)}\big]^{-1} Y_k Q_k^{(2)},
\end{align}
where $\lambda_{j,k}^{(2)}$ is the root of the secular equation $\det\big(\big[\lambda_{k}^{(1)}\big]^{-1} Y_k(z) -\lambda \bm{\text{I}}\big)=0$. Therefore, the transformation from generalized coordinates to canonical coordinates is given as
\begin{align}
X_k(\bm{r})&=\big[Q_{k}^{(2)}\big]^{-1} \big[Q_{k}^{(1)}\big]^{-1}
    \left(\begin{array}{c}
	 u'_{1,k}(\bm{r})\\
	 u_{3,k}(\bm{r})\\
	 \psi_k(\bm{r})
    \end{array}\right)
\nonumber\\
&=U_{0,k}^{\rm c}
\left(\begin{array}{c}
	x_{1,k}(z)\\
	x_{2,k}(z)\\
	x_{3,k}(z)
\end{array}\right) e^{-i k x'} + \text{c.c.},
\label{Canonical-Coordinates-Transformation}
\end{align}
where $U_{0,k}^{\rm c}$ denotes the normalization constant of canonical coordinates, which is determined by the zero-point fluctuation of SAW and will be discussed in Sec.~\ref{SubSecD-Zero-point-Fluctuation}.

Under the transformations $\big[Q_{k}^{(1)}\big]^{-1}$ and $\big[Q_{k}^{(2)}\big]^{-1}$, the kinetic and potential energy terms can be written in the quadratic form. Correspondingly, the Lagrangian is given in canonical coordinate as
\begin{align}
\mathcal{L}=&\sum_{k>0} \mathlarger{\int}
	\frac{1}{2}\rho \bigg(\frac{\partial X_{k}}{\partial t}\bigg)^T
	\lambda_{k}^{\bm{k}}
	\bigg(\frac{\partial X_{k}}{\partial t}\bigg) dV-
	\nonumber\\
	&\sum_{k>0} \mathlarger{\int}
	\frac{1}{2} k^2 X_{k}^T \lambda_{k}^{\bm{p}} X_{k} dV,
\end{align}
where $X_{k}=\big(X_{1,k},\,X_{2,k},\,X_{3,k}\big)^T$~denotes the canonical coordinate. $X_{j,k}=X_{j,k}^{(+)} +X_{j,k}^{(-)}$ is the component of canonical coordinate, with the positive frequency term $X_{j,k}^{(+)} =U_{0,k}^{\rm c} x_{j,k}(z) e^{-i(kx' -\omega_k t)}$ and the negative frequency term $X_{j,k}^{(-)} =U_{0,k}^{\rm c} x_{j,k}(z) e^{i(kx' -\omega_k t)}$. $\lambda_{k}^{\bm{k}}   \equiv {\rm diag}
\big(\lambda_{1,k}^{\bm{k}},\,\lambda_{2,k}^{\bm{k}},\,\lambda_{3,k}^{\bm{k}}\big) =Q_{k}^{(2)T} Y_k Q_{k}^{(2)}$ and $\lambda_{k}^{\bm{p}}   \equiv {\rm diag}\big(\lambda_{1,k}^{\bm{p}},\,\lambda_{2,k}^{\bm{p}},\,\lambda_{3,k}^{\bm{p}}\big) =Q_{k}^{(2)T} Q_{k}^{(1)T} G_{k}(z) Q_{k}^{(1)} Q_{k}^{(2)}$ are the coefficients matrixes. Thus, in terms of the canonical momentum $P_k={\partial \mathcal{L}} \big/ \partial\big({\partial X_k}/{\partial t}\big)$, the Hamiltonian of classical SAW is written as~(see Sec.~III of the Supplemental Material~in detail)
\begin{align}
H_{{\rm saw}}=&\sum_{k>0} P_k \frac{\partial X_k}{\partial t} -\mathcal{L}
    \nonumber\\
   =&\sum_{k>0} \sum_{j=1}^{3}
    \left[
    \lambda_{j,k}^{\bm{k}} \frac{P_{j,k}^2}{2M} +
    \lambda_{j,k}^{\bm{p}} \frac{M k^2}{2\rho} X_{j,k}^2
    \right],
\end{align}
with the mass $M=\int \rho dV$.

\subsection{Quantization condition}
In the above discussion, we have derived the canonical coordinates, canonical momentum and Hamiltonian of classical SAW in the piezomagnetic waveguide. In this subsection, we will introduce the quantization condition to achieve the quantization of SAW. We assume that the canonical coordinates $\hat{X}_{k}$ and the canonical momentum $\hat{P}_{k^{\prime}}$ satisfy the communication relation~$\big[\hat{X}_{k},\hat{P}_{k^{\prime}}\big]=i \hbar \delta_{k k^{\prime}}$. By introducing the creation and annihilation operators $\hat{b}_k^{\dagger} ={\hat{X}_{j,k}}\big/{2X_{j,k}^{(+)}}
-{i\hat{P}_{j,k}}\big/{2M\omega_k \lambda_{j,k}^{\bm{p}} X_{j,k}^{(+)}}$ and $\hat{b}_k ={\hat{X}_{j,k}}\big/{2X_{j,k}^{(-)}}
+{i\hat{P}_{j,k}}\big/{2M\omega_k \lambda_{j,k}^{\bm{p}} X_{j,k}^{(-)}}$, one arrives the canonical coordinate in quantization form as follows
\begin{align}
\hat{X}_k=U_{0,k}^{\rm c}
    \left(\begin{array}{c}
	 x_{1,k}(z)\\
	 x_{2,k}(z)\\
	 x_{3,k}(z)
    \end{array}\right)
\hat{b}_k^{\dagger}  e^{-i(kx' -\omega_k t)}  +\text{H.c.},
\label{Canonical-coordinates-coefficient-quantization-form}
\end{align}
where $\hat{b}_k^{\dagger}$ and $\hat{b}_k$ obey the standard bosonic commutation relations $\big[\hat{b}_k,\hat{b}_{k^{\prime}}^{\dagger}\big]=\delta_{k k^{\prime}}$ and $\big[\hat{b}_k,\hat{b}_{k^{\prime}}\big]=\big[\hat{b}_k^{\dagger},\hat{b}_{k^{\prime}}^{\dagger}\big]=0$. Correspondingly, we have the canonical momentum given in quantization form as
\begin{align}
\hat{P}_k=\mathlarger{\int} i\omega_k \rho U_{0,k}^{\rm c}
    \left(\begin{array}{c}
	  \lambda_{1,k}^{\bm{k}} x_{1,k}(z)\\
	  \lambda_{2,k}^{\bm{k}} x_{2,k}(z)\\
	  \lambda_{3,k}^{\bm{k}} x_{3,k}(z)
    \end{array}\right)
[\hat{b}_k^{\dagger}  e^{-i(kx' -\omega_k t)} -\text{H.c.}] dV.
\end{align}
Thus, the Hamiltonian of SAW is written as
\begin{equation}
\hat{H}_{{\rm saw}}=\sum_{k>0} \hbar \omega_k
\left(
\hat{b}_{k}^{\dagger} \hat{b}_{k} +\frac{1}{2}
\right).
\end{equation}
By taking the inverse transformation of canonical coordinates, i.e.,~$\big(\hat{u}'_{1,k}(\bm{r}),\,\hat{u}_{3,k}(\bm{r}),\,\hat{\psi}_k(\bm{r})\big)^T =Q_{k}^{(1)} Q_{k}^{(2)} \hat{X}_k(\bm{r})$, we can obtain the generalized coordinates given in quantization form as
\begin{align}
\left(\begin{array}{c}
	\hat{u}'_1\\
	\hat{u}_3\\
	\hat{\psi}
\end{array}\right)=\sum_{k>0} U_{0,k}
\left(\begin{array}{c}
	U'_{1,k}(z)\\
	U_{3,k}(z)\\
	\Psi_k(z)
\end{array}\right)
\hat{b}_k e^{i(kx' -\omega_k t)} + \text{H.c.}
\label{Generalized-coordinates-quantization-form}
\end{align}
with the coefficients~$U'_{1,k}(z)$,~$U_{3,k}(z)$,~and~$\Psi_k(z)$~given as
\begin{align}
&U'_{1,k}(z)=2e^{-kq_{\alpha}z} \text{cos}\left(k q_{\beta} z +\theta\right),
\nonumber\\
&U_{3,k}(z)=-2i \left|\gamma\right|e^{-kq_{\alpha}z}\text{cos}\left(k q_{\beta} z +\theta +\xi\right),
\nonumber\\
&\Psi_k(z)=i \frac{q_{33}}{\mu_{11}}
    [2A e^{-k q_{\alpha} z} \text{cos}(k q_{\beta} z +\theta +\tau)+A_3 e^{-kz} ].
\label{Generalized-coordinates-coefficient-quantization-form}
\end{align}
Here, the relation between the normalization constant $U_{0,k}$ of generalized coordinates in Eq.~(\ref{Generalized-coordinates-quantization-form}) and the normalization constant $U_{0,k}^{\rm c}$ of canonical coordinates in Eq.~(\ref{Canonical-coordinates-coefficient-quantization-form}) is determined by the coordinate transformation in Eq.~(\ref{Canonical-Coordinates-Transformation}). It is clear that Eq.~(\ref{Generalized-coordinates-quantization-form}) corresponds to the quantized travelling-wave SAW (propagating along $+x'$ direction) in the waveguide. We note that the SAW resonator can be formed by fabricating two reflecting gratings on the surface of the piezomagnetic medium and the corresponding quantization of SAW in resonators can also be derived in a similar way as for the waveguide by considering the normalization condition.

\begin{figure}[tb]
	\centering
	\includegraphics[scale=0.2]{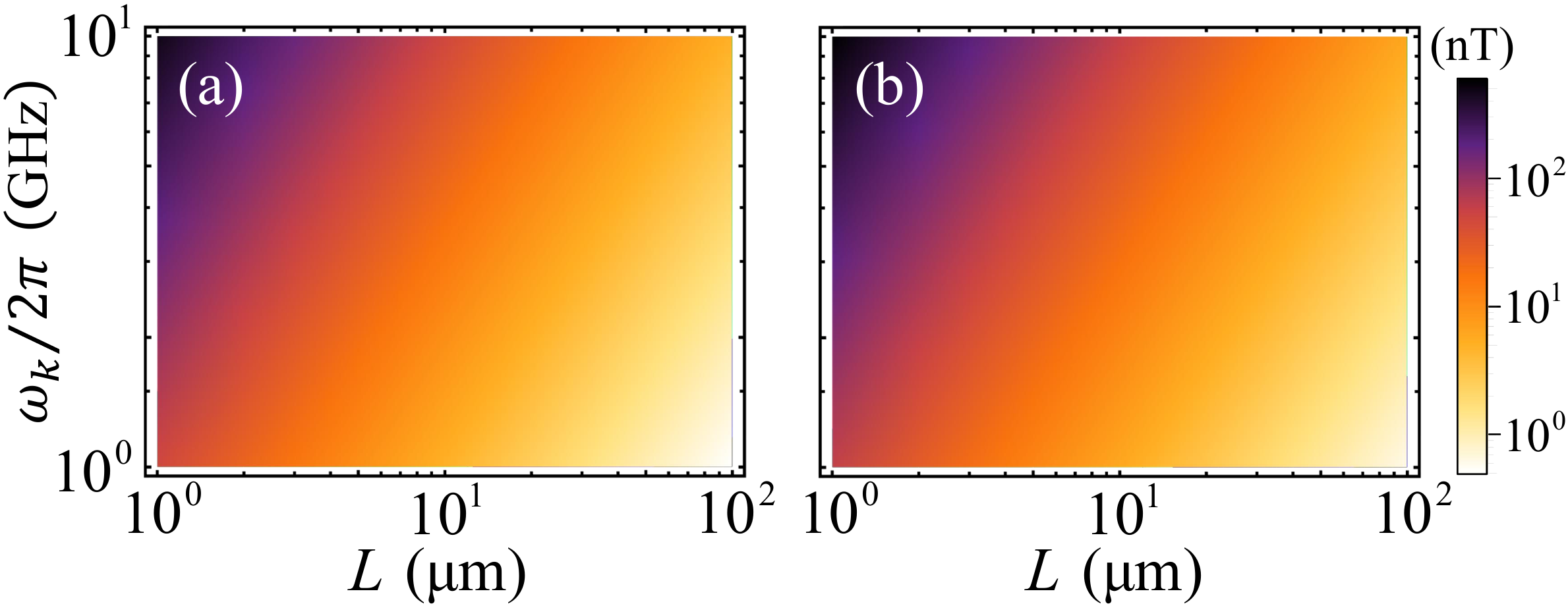}
	\caption{The zero-point fluctuation (a) $B_{x',\,k}^{\mathbf{zp}}$ and (b) $B_{z,\,k}^{\mathbf{zp}}$ of the magnetic field induced by piezomagnetic effect versus the lateral width $L$ and the frequency $\omega_k$ of SAW. Other parameters are the same as those in Fig.~\ref{F2-SAW-xz-dependence}.}
	\label{F3-ZeroPoint-MagneticField}
\end{figure}

\subsection{Zero-point fluctuation of a quantized single-mode SAW}
\label{SubSecD-Zero-point-Fluctuation}
We know that the zero-point fluctuations of a quantized single-mode SAW play an important role in the study of the interaction between quantum systems and quantized single-mode SAW via the magnetic field induced by piezomagnetic effect. In analogy to cavity quantum electrodynamics~\cite{CavityQED-Box-Book}, we here consider a semi-infinite box, with the finite lateral width $L$ in $xy$-plane and the infinite length in $z$ direction. Thus, the normalization constant $U_{0,k}$ of generalized coordinates, which corresponds to the zero-point fluctuation of mechanical displacement and is determined by the energy of a single-mode SAW phonon, can be calculated as $U_{0,k} \simeq (8.71\times 10^{-22} \big/L)\,\text{m}^2$~(see Sec.~IV of the Supplemental Material~in detail). Following the coordinate transformation given in Eq.~(\ref{Canonical-Coordinates-Transformation}) and the generalized coordinates given in Eq.~(\ref{Generalized-coordinates-quantization-form}), one can determine the normalization constant $U_{0,k}^{\rm c}$ of canonical coordinate in Eq.~(\ref{Canonical-Coordinates-Transformation}) accordingly. Here, the expression of $U_{0,k}^{\rm c}$ is not given since it is too complicated. In terms of Eqs.~(\ref{PiezomagneticEq}),~(\ref{Generalized-coordinates-quantization-form}),~and~(\ref{Generalized-coordinates-coefficient-quantization-form}), the components of the induced magnetic field with the wavenumber $k$ on the surface ($z=0$) are written in quantization form as~(see Sec.~IV of the Supplemental Material~in detail)
\begin{align}
\hat{B}_{x',\,k}(z=0)&=
    B_{x',\,k}^{\mathbf{zp}} \hat{b}_{k} e^{i (k x' -\omega_k t)}
    +\text{H.c.},
\label{Bx-quantization-form}
\\%
\hat{B}_{z,\,k}(z=0)&=
    iB_{z,\,k}^{\mathbf{zp}} \hat{b}_{k} e^{i (k x' -\omega_k t)}
    +\text{H.c.},
\label{Bz-quantization-form}
\end{align}
where the corresponding zero-point fluctuations $B_{x',\,k}^{\mathbf{zp}}$~and~$B_{z,\,k}^{\mathbf{zp}}$ are given as
\begin{align}
B_{x',\,k}^{\mathbf{zp}}=&
    2k U_{0,k} q_{33}
    \bigg[A \text{cos}\left(\theta +\tau\right) +\frac{A_3}{2}\bigg],
\label{BxdkZP-detail}
\\%
B_{z,\,k}^{\mathbf{zp}}=  
    &2k U_{0,k} q_{33}
    \big[q_{\alpha} \left|\gamma\right| \text{cos}\left(\theta +\xi\right)
    +q_{\beta} \left|\gamma\right| \text{sin}\left(\theta +\xi\right)\big]+
    \nonumber\\
    &2k U_{0,k} q_{33}
    \big[A q_{\alpha} \text{cos}\left(\theta +\tau\right)
    +A q_{\beta} \text{sin}\left(\theta +\tau\right)\big]+
    \nonumber\\
    &2k U_{0,k} q_{33}
    \bigg[\frac{q_{31}}{q_{33}} \text{cos}\left(\theta\right)
    +\frac{A_3}{2}\bigg].
\label{BzkZP-detail}
\end{align}
In Fig.~\ref{F3-ZeroPoint-MagneticField}, we give the zero-point fluctuations of the induced magnetic fields corresponding to a quantized single-mode SAW with the wavenumber $k$ versus the lateral width $L$ and the frequency $\omega_k$ of SAW. Here, the lateral width $L$ is taken to be $L=1-100\,\SI{}{\micro\meter}$, which is within the width range of the SAW waveguide in the existing researches~\cite{SAW-Waveguide1,SAW-Waveguide2,SAW-Waveguide3}. It is shown that smaller lateral width $L$ and higher frequency $\omega_k$ allow for obtaining the larger zero-point fluctuation of the magnetic field, which would strengthen the coupling between quantum systems and a single-mode SAW via the magnetic field. Specially, we find that the zero-point fluctuation of magnetic fields can reach about $\SI{1\,}{\micro\tesla}$ for $L=\SI{1\,}{\micro\meter}$ and $\omega_k /2\pi =10\,\text{GHz}$. In such a case, for the mechanical displacement which differs from the magnetic field by a factor $k q_{ijk}$, its corresponding zero-point fluctuation is about $1\,\text{fm}$.

\section{Coupling between magnetic quantum systems and a quantized single-mode SAW in piezomagnetic waveguide }
\label{Sec4-SAW-Coupling-to-Systems}

In above discussions, we have obtained the quantized SAW in the waveguide made of piezomagnetic medium. Different from the SAW in the piezoelectric medium, the SAW in the piezomagnetic medium allows direct manipulations of magnetic quantum systems via the magnetic field induced by the piezomagnetic effect. SAW is usually designed to operate in the frequency range from tens of megahertz to several gigahertz. We note that the superconducting circuit, ferromagnetic magnon, and defect center in diamond are typical solid-state magnetic quantum systems (with magnetic moments), which can operate in the domain of microwave frequency. Without loss of generality, we now study the coupling between these quantum systems and a quantized single-mode SAW at single phonon level (i.e., the coupling between these quantum systems and a single-mode SAW phonon) in the piezomagnetic waveguide.

\begin{figure*}[tb]
	\centering
	\includegraphics[width=13cm]{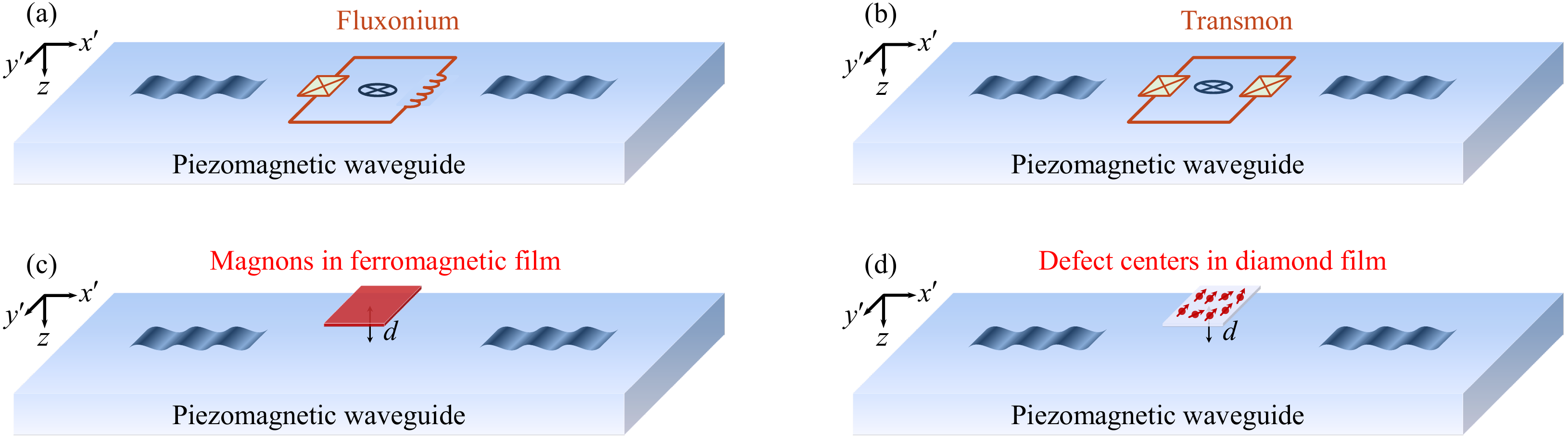}
	\caption{Schematic illustration for coupling the (a) fluxonium qubit, (b) transmon qubit, (c) ferromagnetic magnon, and (d) defect center in diamond with quantized SAW in one-dimensional piezomagnetic waveguide via the magnetic field induced by piezomagnetic effect. The superconducting qubits are deposited on the surface of the waveguide, while the ferromagnetic film and diamond film are suspended above the waveguide.}
	\label{F4-SAW-MagneticSystems-Coupling}
\end{figure*}

\subsection{Superconducting circuits}
The superconducting qubit circuits based on Josephson junctions are macroscopic quantum systems, which have the advantages such as the flexibly designed and tunable system parameters, long decoherence time, as well as capability of the interface with other quantum systems~\cite{Superconducting-Electromechanics,Superconducting-Circuit1,Superconducting-Circuit2,
Superconducting-Circuit3,Superconducting-Circuit4}. It is well known that the superconducting qubit loop interrupted by Josephson junctions is very sensitive to the magnetic field~\cite{SQUID-Original,SQUID-Analysis,SQUID-Tunable-Coupling,SQUID-Book}. We here take the fluxonium~\cite{Fluxonium-Original,Fluxonium1,Fluxonium2,Fluxonium3} and transmon~\cite{Transmon-Original,Transmon1,Transmon2,Transmon3}~qubits as examples to study their couplings to a quantized single-mode SAW in the piezomagnetic waveguide via the magnetic field induced by the piezomagnetic effect. In the following study, we simply assume that the effect of the magnetic field induced by SAW on the superconductivity of the superconducting qubit is negligibly small. And also, the superconducitng qubit might be coupled to the SAW in the piezomagnetic waveguide through flip-chip bonding technology~\cite{SAW-Entanglement1,Flip-Chip-Bonding1,Flip-Chip-Bonding2} to minimize the effect of the piezomagnetic waveguide on the superconductivity and coherence of the qubit.

\subsubsection{Coupling of a quantized single-mode SAW to fluxonium qubit}
We first study the coupling between the fluxonium qubit and a quantized single-mode SAW in the piezomagnetic waveguide. As schematically shown in Fig.~\ref{F4-SAW-MagneticSystems-Coupling}(a), the fluxonium qubit, which contains a superconducting loop interrupted by a Josephson junction and a big linear inductor, is coupled to SAW via the magnetic field induced by piezomagnetic effect. Thus, the Hamiltonian of the system~(see Sec.~V of the Supplemental Material~in detail), that the fluxonium qubit is coupled to a quantized single-mode SAW in the piezomagnetic waveguide, is written as~(Hereafter, we take $\hbar=1$ for simplicity.)
\begin{align}
H_{fx}=\,& \omega_{fx}\,\hat{a}_{fx}^{\dagger} \hat{a}_{fx}
	-\frac{E_c}{12}\left(\hat{a}_{fx}+\hat{a}_{fx}^{\dagger}\right)^4+
	\omega_k \hat{b}_{k}^{\dagger}\hat{b}_{k}+
	\nonumber\\
	&\big(i g_{fx} \hat{a}_{fx}^{\dagger} \hat{b}_k +\text{H.c.}\big),
\label{SAW-FluxoniumQubit-Total-Hamiltonian}
\end{align}
where $\omega_{fx}=2\sqrt{2E_C E_J} +E_L \sqrt{2E_C / E_J}$ is the plasma-like frequency. $\hat{a}_{fx}^{\dagger}$~and~$\hat{a}_{fx}$ are the creation and annihilation operators of the fluxonuim qubit. $g_{fx}=-{2\pi}\sqrt[4]{2E_C / E_J} E_L B_{z,\,k}^{\mathbf{zp}} S \big/{\Phi_0}$ is the coupling strength between the fluxonium qubit and a quantized single-mode SAW via the magnetic field induced by piezomagnetic effect. $E_C$, $E_J$, and $E_L$ denote the capacitive, Josephson, and inductive energies of the fluxonium qubit, respectively. $\Phi_0$ is magnetic flux quantum. $B_{z,\,k}^{\mathbf{zp}}$ denotes the zero-point fluctuation of the induced magnetic field along $z$ direction as given in Eq.~(\ref{BzkZP-detail}). $S$ denotes the effective loop area of the qubit that the magnetic field threads. Equation~(\ref{SAW-FluxoniumQubit-Total-Hamiltonian}) shows that the fluxonium qubit is linearly coupled to a quantized single-mode SAW in the waveguide via the magnetic field induced by the piezomagnetic effect.

\subsubsection{Coupling of a quantized single-mode SAW to transmon qubit}
We now study the coupling between the transmon qubit and a quantized single-mode SAW in the piezomagnetic waveguide. As schematically shown in Fig.~\ref{F4-SAW-MagneticSystems-Coupling}(b), the transmon qubit, which contains a superconducting loop interrupted by two Josephson junctions, is coupled to SAW via the magnetic field induced by piezomagnetic effect. Thus, the Hamiltonian of the system~(see Sec.~VI of the Supplemental Material~in detail), that the transmon qubit is coupled to a quantized single-mode SAW in the piezomagnetic waveguide, is given as
\begin{align}
H_{ts}=\,& \omega_{ts}\,\hat{a}_{ts}^{\dagger} \hat{a}_{ts}
    -\frac{E_c}{12}\left(\hat{a}_{ts}+\hat{a}_{ts}^{\dagger}\right)^4
    +\omega_k \hat{b}_{k}^{\dagger}\hat{b}_{k}+
    \nonumber\\
    &g_{ts} (\hat{a}_{ts}+\hat{a}_{ts}^{\dagger})^2
    \big(\hat{b}_{k} -\hat{b}_{k}^{\dagger}\big)^2,
\label{SAW-TransmonQubit-Total-Hamiltonian}
\end{align}
where $\omega_{ts}=4\sqrt{E_C E_J}$ denotes the Josephson plasma frequency. $\hat{a}_{ts}^{\dagger}$~and~$\hat{a}_{ts}$ are the creation and annihilation operators of the transmon qubit. The parameter $g_{ts}=-{\sqrt{E_C E_J} (\pi B_{z,\,k}^{\mathbf{zp}} S)^2} \big/ {2\Phi_0^2}$ is the coupling strength between the transmon qubit and a quantized single-mode SAW via the magnetic field induced by piezomagnetic effect. $E_C$, $E_J$, and $E_L$ denot the capacitive, Josrphson, and inductive energies of the transmon qubit, respectively. Equation~(\ref{SAW-TransmonQubit-Total-Hamiltonian}) shows that different from the fluxonium qubit, the transmon qubit is quadratically coupled to a quantized single-mode SAW via the magnetic field induced by piezomagnetic effect.

\begin{figure}[t]
	\centering
	\includegraphics[scale=0.2]{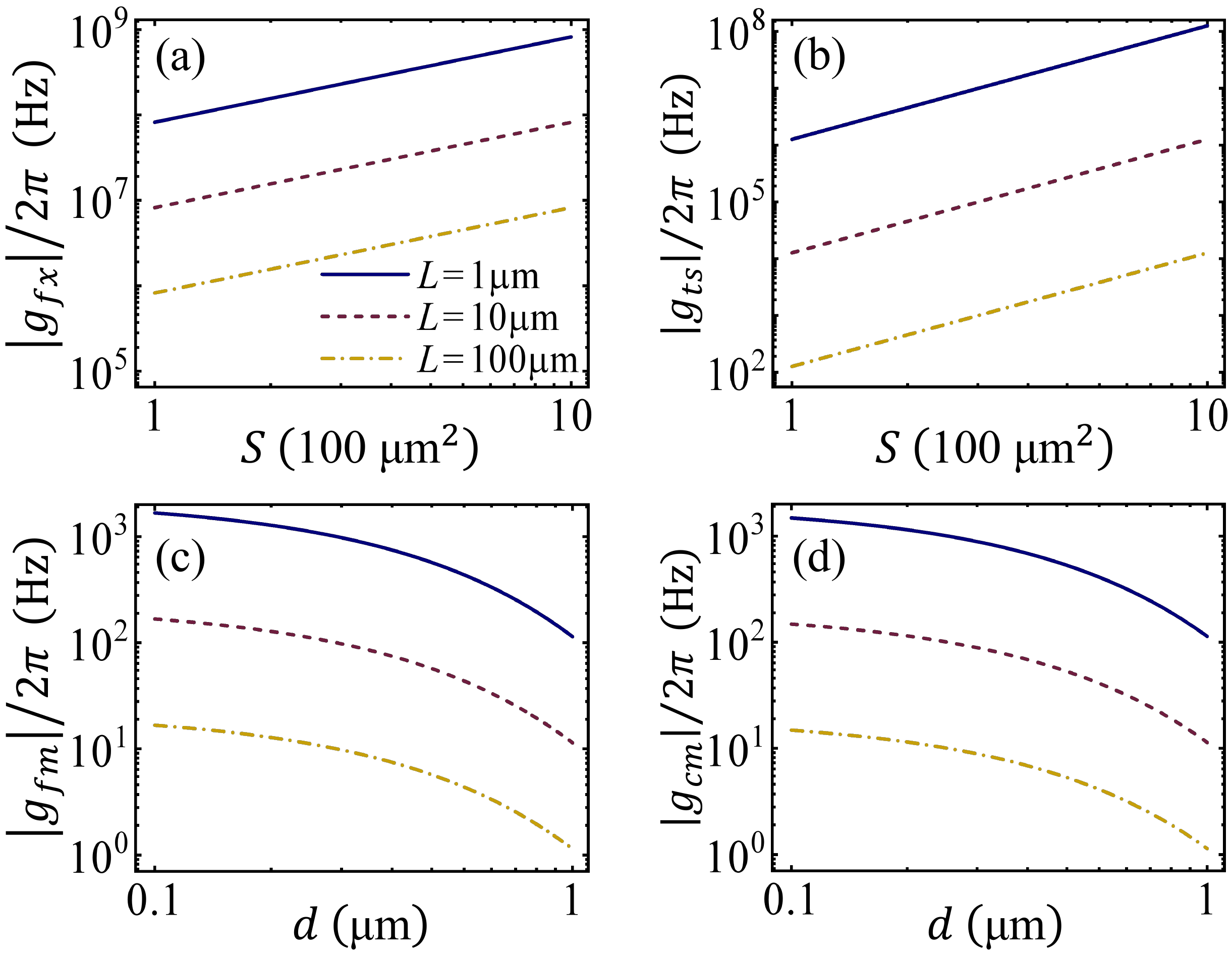}
	\caption{The coupling strength between the (a) fluxonium qubit (b) transmon qubit and a quantized single-mode SAW via the magnetic field induced by piezomagnetic effect versus the loop area $S$ for different lateral width $L$. The coupling strength between single (c) ferromagnetic magnon (d) defect center and a quantized single-mode SAW via the magnetic field induced by piezomagnetic effect versus the distance $d$ for different lateral width $L$. The parameters are chosen as: (a) $E_C/2\pi =E_L/2\pi =1\,\text{GHz}$, $E_J/2\pi =3\,\text{GHz}$, and $\omega_k /2\pi=4.72\,\text{GHz}$; (b) $E_C/2\pi =100\,\text{MHz}$, $E_J/2\pi =10\,\text{GHz}$, and $\omega_k /2\pi=3.9\,\text{GHz}$; (c) $\gamma_f /2\pi=30.59\,\text{GHz}$ and $\omega_k /2\pi=3\,\text{GHz}$; (d) $\gamma_c /2\pi=28\,\text{GHz}$ and $\omega_k /2\pi=2.87\,\text{GHz}$. Other parameters are the same as those in Fig.~\ref{F2-SAW-xz-dependence}.}
	\label{F5-SAW-Qubits-MagneticCouplingStrength}
\end{figure}

\subsubsection{Coupling strength between a quantized single-mode SAW and superconducting qubits}
In Figs.~\ref{F5-SAW-Qubits-MagneticCouplingStrength}(a)~and~\ref{F5-SAW-Qubits-MagneticCouplingStrength}(b), we choose different lateral width $L$ to display the coupling strengths in Eqs.~(\ref{SAW-FluxoniumQubit-Total-Hamiltonian})~and~(\ref{SAW-TransmonQubit-Total-Hamiltonian}) of the fluxonium and the transmon qubits with a quantized single-mode SAW versus the loop area $S$ of the qubit. Here, the frequency of SAW is taken to be resonant with the transition frequency between two lowest energy states of the fluxonium and transmon qubits in Figs.~\ref{F5-SAW-Qubits-MagneticCouplingStrength}(a)~and~\ref{F5-SAW-Qubits-MagneticCouplingStrength}(b), respectively. It is found that with the decrease of $L$ or the increase of $S$, the coupling strength can be enhanced significantly. Therefore, in the case of superconducting qubits that are coupled to a single-mode SAW via the magnetic field, designing phononic devices with small lateral widths or qubit circuits with large loop areas would be necessary to achieve the strong coupling strength.

Furthermore, we point out that besides the linear coupling, a single-mode SAW in piezomagnetic waveguide can interact with superconducting qubit via the quadratic coupling. This is different from the quantized single-mode SAW in piezoelectric waveguide, which typically interacts with a superconducting qubit via the linear coupling. In the present models, the linear coupling or quadratic coupling arises from the intrinsic properties of the qubits. Thus, when resorting to special-designed superconducting qubit circuit, one could enable the interaction between the qubits and SAW via other types of nonlinear couplings. This would bring more potential applications in quantum communication and quantum computing.

\subsection{Coupling of a quantized single-mode SAW to ferromagnetic magnons}

The magnon~\cite{Ferrimagnet-MWcavity-Original,Ferrimagnet-Review} is the spin wave quantum in magnetically ordered systems. Ferromagnet~\cite{Ferrimagnet-MWcavity-Original,Ferrimagnet-MWcavity,Ferrimagnet-Experiment,Ferrimagnet-Applications,Ferrimagnet-Review,
Ferrimagnet-Film-Superconducting,Ferrimagnet-QuantumRegime,Ferrimagnet-Film-Decoherence,Ferrimagnet-Nickel-SAW,Ferrimagnet-Sphere-Transducer}, which is a typical example of magnetically ordered systems, is well known for its rich properties, e.g., large magnetization, long decoherence time, magnetostriction, and nonlinearity. These properties allow ferromagnetic magnons for promising applications in quantum communication, quantum sensor, and quantum transduction~\cite{Ferrimagnet-Applications,Ferrimagnet-Review,SAW-Magnetic-Film1,SAW-Magnetic-Film2,SAW-Magnetic-Film3,SAW-Magnetic-Film4}. Here, we study the interaction between ferromagnetic magnons and a quantized single-mode SAW in piezomagnetic waveguide.

Note that the magnetic potential of SAW decays exponentially in the space above the waveguide made of piezomagnetic medium. To achieve the coupling between the ferromagnetic magnon and a quantized single-mode SAW in waveguide via the magnetic field induced by piezomagnetic effect, we consider the system composed of the piezomagnetic waveguide and ferromagneitc film suspended above the waveguide as schematically shown in Fig.~\ref{F4-SAW-MagneticSystems-Coupling}(c). The corresponding interaction Hamiltonian is given as~(see Sec.~VII of the Supplemental Material~in detail)
\begin{align}
H_{fm}&=g_{fm} \hat{m}^{\dagger} \hat{b}_k +\text{H.c.},
\end{align}
where $g_{fm}=(1-i)\sqrt{Ns}\gamma_{f} B_{x',k}^{\mathbf{zp}} e^{-kd} \big/2$ denotes the coupling strength between $N$ ferromagnetic spins and a quantized single-mode SAW. $s$ is the spin quantum number. $\gamma_{f}$ is the gyromagnetic ratio. $d$ is the distance between the film and the surface of waveguide. $\hat{m}^{\dagger}$ and $\hat{m}$ are the creation and annihilation operators of the magnon. As shown in Fig.~\ref{F5-SAW-Qubits-MagneticCouplingStrength}(c), for the typical ferromagnetic material Nickel with operation frequency~$\omega_{f} /2\pi=3\,\text{GHz}$, gyromagnetic ratio~$\gamma_{f}/2\pi \simeq 30.59\,\text{GHz}$, and spin quantum number $s=1/2$~\cite{Ferrimagnet-Nickel-SAW}, when we take the distance $d=0.1\,\SI{}{\micro\meter}$ and the lateral width $L=1\,\SI{}{\micro\meter}$, the coupling strength between single ferromagnetic magnon and single-mode SAW phonon with frequency $\omega_k =\omega_{f}$ is estimated as $g_{fm} /2\pi \simeq 1673\,\text{Hz}$. The results indicate that, smaller lateral width $L$ for the SAW device and the smaller distance $d$ between the SAW device and the  ferromagnetic film would contribute to realize strong coupling strength.

Additionally, we point out that when the ferromagnetic film is deposited on the surface of elastic medium, the propagation of SAW through the film would generate elastic strain, allowing to drive the magnetization dynamics of the ferromagnetic magnons via the magnetoelastic interaction~\cite{SAW-Magnetic-Film1,SAW-Magnetic-Film2,SAW-Magnetic-Film3,SAW-Magnetic-Film4}. Here, it is found that when the ferromagnetic magnons are coupled to SAW via the strain induced by piezomagnetic effect, one can achieve the interaction between ferromagnetic magnons and SAW via both linear, nonlinear, and magnomechanical couplings simultaneously~(see Sec.~VII of the Supplemental Material~in detail). This would offer promising applications in quantum entanglement.

\begin{table*}[t]
\centering
\caption{Comparisons of coupling strengths between typical magnetic quantum systems and a quantized single-mode SAW in the piezomagnetic waveguide. Other parameters are $L=1-100\,\SI{}{\micro\meter}$, $S=100-1000\,\SI{}{\micro\meter}$$^2$, and $d=0.1-1\,\SI{}{\micro\meter}$.}
\renewcommand{\arraystretch}{1.8}
\begin{tabular}
{m{5cm}<{\centering}|p{2.8cm}<{\centering}|p{2.8cm}<{\centering}|p{3.2cm}<{\centering}|p{3.2cm}<{\centering}}
		\hline
	\bf{Qubits}
	&\bf{Fluxonium}
	&\bf{Transmon}
	&\bf{Nickel for magnon}
	&\bf{NV center for defect center} \\
		\hline
	Working temperature
	&$\sim$10\,mK
	&$\sim$10\,mK
	&from $\sim$1\,K to room temperature
	&room temperature \\
		\hline
	Working frequency $\omega \big/ 2\pi$
	&$\sim1-5$\,GHz
	&$\sim1-5$\,GHz
	&$\sim0.1-5$\,GHz
	&$\sim1-5$\,GHz \\
		\hline
	Coupling type
	&either electric or magnetic
	&either electric or magnetic
	&either magnetic or strain
	&either electric, magnetic, or strain\\
		\hline
	Decoherence rate $\Gamma \big/ 2\pi$ of single qubit
	&$\sim 10^{-1}$\,MHz
	&$\sim 10^{-1}$\,MHz
	&$\sim 1-10$\,kHz
	&$\sim 1\,\text{Hz}-1$\,kHz\\
		\hline
	Coupling strength $g \big/ 2\pi$ between single qubit and SAW (in waveguide) via magnetic field
	&$\sim 10^{-1}-10^2$\,MHz
	&$\sim 1\,\text{Hz}-10^2$\,MHz
	&$\sim 1\,\text{Hz}-1$\,kHz
	&$\sim 1\,\text{Hz}-1$\,kHz\\
		\hline
\end{tabular}
\label{Coupling-Strength-SolidStateSystems}
\end{table*}

\subsection{Coupling of a quantized single-mode SAW to defect centers in diamond}

The defect center in diamond, which is considered as a spin, can operate at room temperature and has long decoherence time. Thus, it provides a promising platform for quantum memory, quantum communication, and quantum sensor~\cite{NVcenter-Original,SAW-NVcenter-Coupling1,NVcenter-Progress,NVcenter-PESAW-Device,SAW-NVcenter-Coupling2,NVcenter-PM-Lattice,NVcenter-Sensor-magnetic-strain,NVcenter-EMfield-Telecom1,NVcenter-EMfield-Telecom2,NVcenter-Mfield-Memory}. Different from the previous studies involving the interaction between the electromagnetic fields at telecommunication wavelength~\cite{NVcenter-EMfield-Telecom1,NVcenter-EMfield-Telecom2,NVcenter-Mfield-Memory}~or SAW in piezoelectric medium~\cite{NVcenter-PESAW-Device,SAW-NVcenter-Coupling1,SAW-NVcenter-Coupling2}, we here focus on the domain of microwave frequency to study the coupling between the defect center and SAW in piezomagnetic medium.

Similarly, the coupling between defect centers and SAW can be realized via the magnetic field induced by piezomagnetic effect. As schematically shown in Fig.~\ref{F4-SAW-MagneticSystems-Coupling}(d), we consider the system composed of the piezomagnetic waveguide and defect centers contained in the diamond film, which is suspended above the waveguide. Correspondingly, the interaction Hamiltonian is written as~(see Sec.~VIII of the Supplemental Material~in detail)
\begin{align}
H_{cm}=g_{cm} \sigma_{+} \hat{b}_k +\text{H.c.},
\end{align}
where $g_{cm}=(1-i) \gamma_{c} B_{x',\,k}^{\mathbf{zp}} e^{-k d} \big/2\sqrt{2}$ is the coupling strength between single defect center and a quantized single-mode SAW, with the gyromagnetic ratio of defect center $\gamma_c$. $\sigma_{+}$ and $\sigma_{-}$ are the ladder operators of defect center. Typically, for the NV center with transition frequency~$\omega_{c} /2\pi=2.87\,\text{GHz}$ and gyromagnetic ratio~$\gamma_c /2\pi \simeq 28\,\text{GHz}$~\cite{NVcenter-Progress}, when we take $L=1\,\SI{}{\micro\meter}$ and $d=0.1\,\SI{}{\micro\meter}$, the coupling strength between single NV center and single-mode SAW phonon with frequency $\omega_k =\omega_{c}$ is about $g_{cm} /2\pi \simeq 1484\,\text{Hz}$ as shown in Fig.~\ref{F5-SAW-Qubits-MagneticCouplingStrength}(d). Similar to the case of the coupling between the ferromagnetic magnon and SAW, smaller lateral width $L$ for the SAW device and the smaller distance $d$ between the SAW device and the defect center would be better to obtain strong coupling strength.

Moreover, we note that besides the electric and magnetic fields, the defect centers can be controlled via strain using electromechanical or optomechanical systems~\cite{NVcenter-Strain-Electric-Theory,NVcenter-in-MechanicalCavity,NVcenter-in-Cantilever1,
NVcenter-in-NanoOptomechanics1,NVcenter-in-NanoOptomechanics2}. Thus, in our present model, when the diamond film is deposited on the surface of piezomagnetic waveguide, one can manipulate the defect centers using SAW via both the magnetic field and strain induced by piezomagnetic effect~(see Sec.~VIII of the Supplemental Material~in detail). Specially, for the typical strain-driven NV center~\cite{NVcenter-Strain-Electric-Theory}, its maximum coupling strength with SAW via the strain is estimated as $g_{cs} /2\pi\sim 100\,\text{Hz}$ for the lateral width $L=\SI{1\,}{\micro\meter}$. That means, the coupling strength of NV center with strain is much smaller than it coupling strength with magnetic field, which is in agreement with the previous researches~\cite{QuantumAcoustics-Article1-110-PESAW-Quantization,NVcenter-Sensor-magnetic-strain}.

\subsection{Discussions about coupling strengths}

In the above studies, we have studied the coupling between the solid-state magnetic quantum systems (i.e., superconducting qubit, ferromagnetic magnon, and defect center in diamond) and a single-mode SAW phonon in the waveguide made of piezomagnetic medium. We note that for these systems used as qubits, each of them has its own features, and their couplings to the SAW phonon are also different from each other. For comparisons, we summarize the operating temperature, operating frequency, coupling type, coupling strength to the SAW phonon via the magnetic field, and decoherence rate of these systems in Table~\ref{Coupling-Strength-SolidStateSystems}.

The superconducting qubit, which operates at low temperature ($\sim10\,\text{mK}$), can interact strongly with a SAW phonon since the coupling strength ($\sim 1\,\text{Hz}-10^2$\,MHz) can be stronger compared to the decoherence rate ($\sim 10^{-1}$\,MHz). Thus, SAW in the piezomagnetic medium offers an alternative way for the individual manipulation of superconducting qubits via the magnetic field induced by piezomagnetic effect. Differently, compared with the superconducting qubit, the ferromagnetic magnon and defect center have smaller decoherence rate ($\sim 1\,\text{Hz}-10$\,kHz) and operates at higher temperature ($\sim1\,\text{K} - 300\,\text{K}$). However, it is difficult to achieve the strong interaction between a single ferromagnetic magnon (or single defect center) and a SAW phonon, since the coupling strength ($\sim 1\,\text{Hz}-1$\,kHz) is comparable or much smaller than the decoherence rate. Thus, a sample of ferromagnetic materials with high spin density~\cite{Ferrimagnet-Nickel-SAW,SAW-Magnetic-Film4,Ferrimagnet-Film-Superconducting} (or diamond with many defect centers~\cite{SAW-NVcenter-Coupling2,NVcenter-in-Cantilever1,NVcenter-in-ThinFilm1}) is required to achieve the coherent manipulation of the collective spins, since the coupling strength between collective spins and a single-mode SAW phonon can be enhanced collectively. Moreover, besides the magnetic field, the ferromagnetic magnons (or defect centers in diamond) can interact with SAW via the strain induced by piezomagnetic effect. Therefore, SAW in the piezomagnetic medium provides an alternative way for the manipulation of ferromagnetic magnons or defect centers in diamond via the magnetic field or strain induced by piezomagnetic effect.

We point out that the above studies focus on the case of the travelling-wave SAW in waveguide, whose amplitude is uniform in space. Thus, the corresponding coupling strength is independent on the locations of qubits. However, in the case of the standing-wave SAW in resonator which will be studied in our future work, the amplitude of SAW is space-dependent, and thus the coupling strengths of the magnetic quantum systems to SAW are also space-dependent. We also note that the wavelength of SAW ($\sim1\,\SI{}{\micro\meter}$ in our present studies) is comparable to the size of single superconducting qubit ($\sim1-100\,\SI{}{\micro\meter}$)~\cite{Superconducting-Circuit1,Superconducting-Circuit2,Superconducting-Circuit3,Superconducting-Circuit4} or a sample~($\sim10-100\,\SI{}{\micro\meter}$) containing many ferromagnetic magnons~\cite{Ferrimagnet-Nickel-SAW,SAW-Magnetic-Film4,Ferrimagnet-Film-Superconducting} or defect centers~\cite{SAW-NVcenter-Coupling2,NVcenter-in-Cantilever1,NVcenter-in-ThinFilm1}. Therefore, these magnetic quantum systems cannot be considered as point-particles when they are coupled to SAW in piezomagnetic resonator.

\section{Qubits interaction mediated by multi-mode SAW phonon in piezomagnetic waveguide}
\label{Sec5-SAW-mediated-Interaction}

In the last section, we have studied the couplings between several typical magnetic quantum systems and a single-mode SAW phonon in piezomagnetic waveguide. Actually, when a magnetic quantum system is coupled to the SAW in waveguide, it would simultaneously interact with multiple modes of SAW phonons in waveguide. In this section, by taking the superconducting fluxonium qubit as an example, we study the distant interaction between two identical qubits mediated by multi-mode SAW phonons in piezomagnetic waveguide. The transition frequencies $\omega_{10}$ of two qubits are taken as $\omega_{10}\big/2\pi=4.72\,\text{GHz}$ according to parameters of Fig.~\ref{F5-SAW-Qubits-MagneticCouplingStrength}(b) in the following study. To characterize such interaction, we focus on the dynamical evolutions of population and entanglement of the two qubits.

\subsection{Dynamical equations of population and concurrence}
\begin{figure}[t]
	\centering
	\includegraphics[scale=0.26]{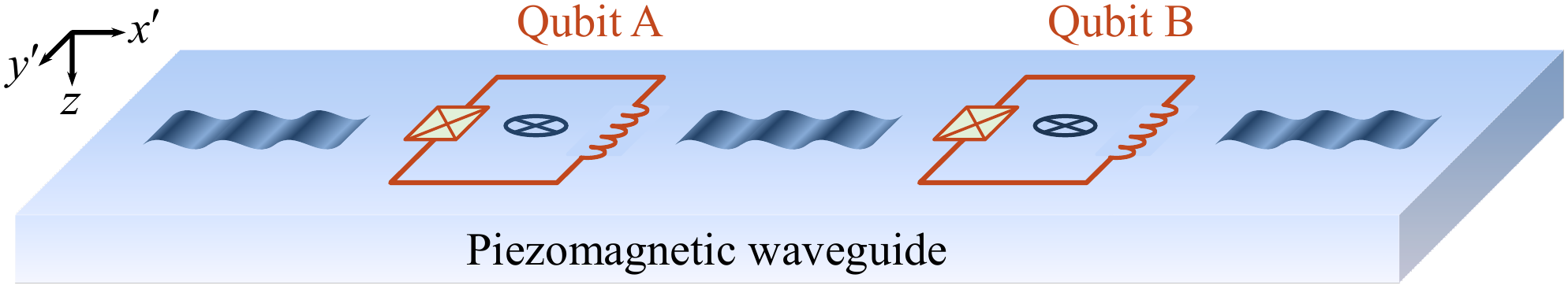}
	\caption{Schematic illustration for coupling two distant superconducting fluxonium qubits mediated by multi-mode SAW phonons in one-dimensional waveguide via the magnetic field induced by piezomagnetic effect.}
	\label{F6-Qubits-Interaction-Via-SAW}
\end{figure}

As schematically shown in Fig.~\ref{F6-Qubits-Interaction-Via-SAW}, the SAW phonons in one-dimensional waveguide are coupled to two superconducting fluxonium qubits at positions $x'=x'_a$ and $x'=x'_b$, respectively. We here focus on the single-excitation properties of the system, thus the evolutions of the two qubits are given as (see Sec.~IX of the Supplemental Material~in detail)
\begin{align}
\frac{\partial}{\partial t} \alpha_A(t)=&
	-\Gamma_0  \alpha_A(t)
	-\Gamma_0  e^{i \theta_{\mathbb{T}}}
	\alpha_B(t-\mathbb{T})  \Theta\left(t-\mathbb{T}\right),
\label{A-Atom-dynamics}
\\%
\frac{\partial}{\partial t} \alpha_B(t)=&
	-\Gamma_0  \alpha_B(t)
	-\Gamma_0  e^{i \theta_{\mathbb{T}}}
	\alpha_A(t-\mathbb{T})  \Theta\left(t-\mathbb{T}\right).
\label{B-Atom-dynamics}
\end{align}
where $\alpha_A(t)$ and $\alpha_B(t) $ are the probability amplitudes of the qubit A and qubit B, which are excited to their first-excited states $|1\rangle_A$ and $|1\rangle_B$ by the SAW phonons. The parameter $\theta_{\mathbb{T}}=\omega_{10} \mathbb{T}$ represents the phase that results from the propagation of SAW, with the transition frequency~$\omega_{10}$ between two lowest energy states of each qubit and the delay time $\mathbb{T}= \left|x'_b -x'_a\right| /v$ between two qubits. $\Theta(t-\mathbb{T})$ is the Heaviside step function. $\Gamma_0$ is the decay rate from the qubit to the SAW waveguide, while other decay rates of the qubit have been neglected. Thus, the population evolutions of the two qubits are calculated as $P_A(t)=\left|\alpha_A(t)\right|^2$~and~$P_B(t)=\left|\alpha_B(t)\right|^2$. Also,
equations~(\ref{A-Atom-dynamics})~and~(\ref{B-Atom-dynamics}) show that the SAW waveguide results in decays of two qubits and induces the interaction between two qubits.

To quantify the entanglement of two qubits, we here resort to the concurrence~\cite{Entangle-Defination-Original,Entangle-Defination1-Book}, derived from the reduced density matrix in the basis~$\{\left|1_A,\,1_B\right\rangle,\,\left|1_A,\,0_B\right\rangle,\,\left|0_A,\,1_B\right\rangle,\,\left|0_A,\,0_B\right\rangle\}$. Thus, the time evolution of the concurrence $C(t)$ is given as~(see Sec.~IX of the Supplemental Material~in detail)
\begin{align}
C(t)&=2\big|\alpha_A(t) \alpha_B(t)^*\big|,
\end{align}
where the probability amplitudes $\alpha_A(t)$~and~$\alpha^*_B(t)$ are obtained by solving the dynamical equations in Eqs.~(\ref{A-Atom-dynamics})~and~(\ref{B-Atom-dynamics}).

\begin{figure}[tbp]
	\centering
	\includegraphics[scale=0.21]{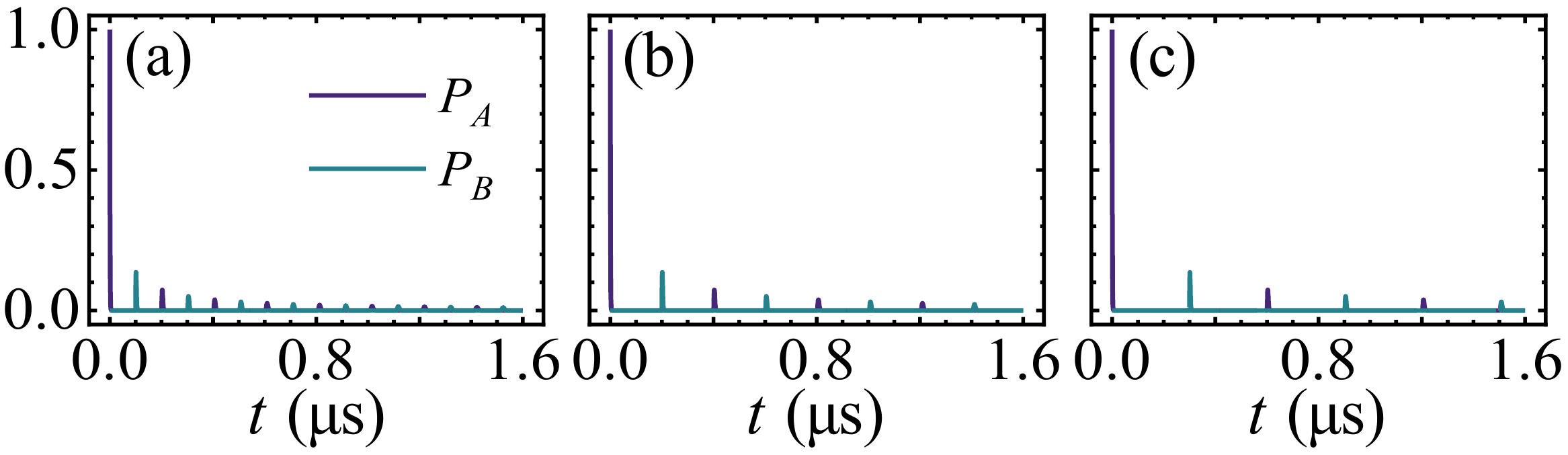}
	\caption{The population evolution of the two qubits when the delay time is taken as (a) $\mathbb{T}=\SI{0.1\,}{\micro\second}$, (b) $\mathbb{T}=\SI{0.2\,}{\micro\second}$, and (c) $\mathbb{T}=\SI{0.3\,}{\micro\second}$. Other parameters are $\omega_{10}\big/2\pi=4.72\,\text{GHz}$ and $\Gamma_0 /2\pi=100\,\text{MHz}$.}
	\label{F7-Qubits-Dynamics-via-SAW-Delay}
\end{figure}

\subsection{Time-delay interaction of two qubits}
In the present model, the time-delay effect due to the propagation of SAW in the waveguide is essential in the interaction between the two qubits. Thus, we first choose different delay times $\mathbb{T}$ to study the population evolutions of two qubits in Fig.~\ref{F7-Qubits-Dynamics-via-SAW-Delay}. In the following discussions, we consider the initial condition that the qubit A (B) is in the excited (ground) state. As shown in Fig.~\ref{F7-Qubits-Dynamics-via-SAW-Delay}(a), at the time $0 \leq t < \mathbb{T}$, the qubit A decays exponentially, while the qubit B remains in the ground state. This is because the SAW phonons, generated by the decay of the qubit A, have not yet propagated to the position of the qubit B. At the time $\mathbb{T} \leq t < 2\mathbb{T}$, the qubit B is excited and then decays exponentially, while the qubit A remains in the ground state. This can be interpreted as, the qubit B is excited by the SAW phonons generated by the qubit A, however the SAW phonons generated by the decay of the qubit B have not yet propagated to the qubit A due to the delay time. As time $t$ goes on, more energy is transferred to the SAW phonons in waveguide, which ultimately leads to the decay of time-delay Rabi oscillation between two qubits. Here, we note that with the increase of delay time $\mathbb{T}$, the time required for SAW phonons propagating between the two qubits is increased accordingly, and thus the decay of Rabi oscillation becomes slower as shown in  Figs.~\ref{F7-Qubits-Dynamics-via-SAW-Delay}(b)~and~\ref{F7-Qubits-Dynamics-via-SAW-Delay}(c).

\begin{figure}[tbp]
	\centering
	\includegraphics[scale=0.21]{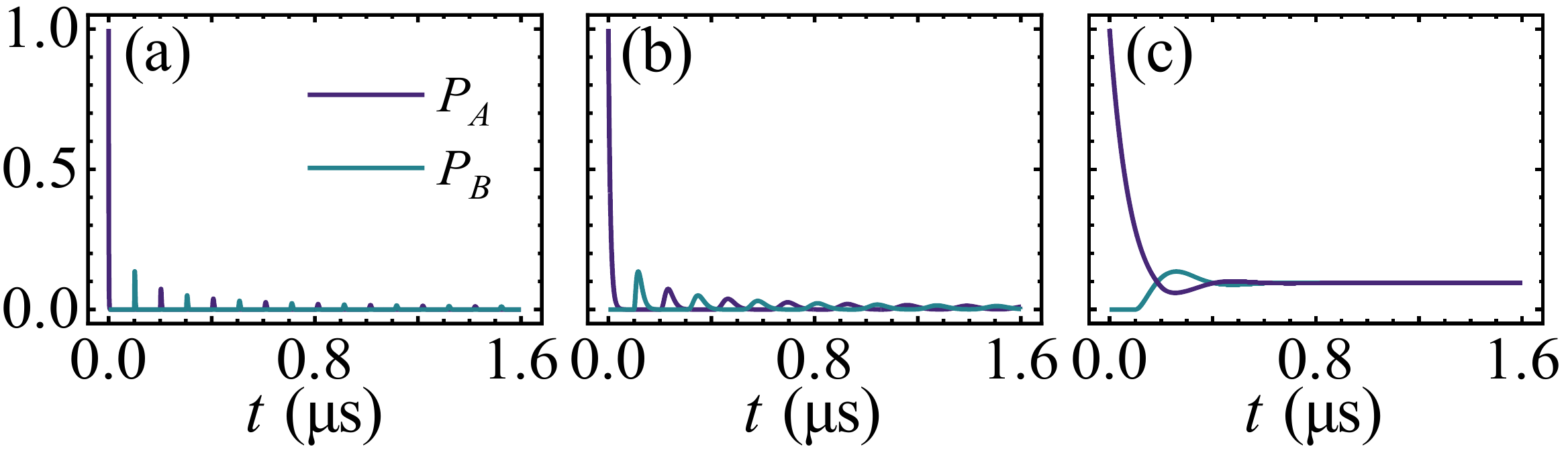}
	\caption{The population evolution of the two qubits when the decay rate to SAW waveguide is taken as (a) $\Gamma_0 /2\pi=100\,\text{MHz}$, (b) $\Gamma_0 /2\pi=10\,\text{MHz}$, and (c) $\Gamma_0 /2\pi=1\,\text{MHz}$. Other parameters are the same as those in Fig.~\ref{F7-Qubits-Dynamics-via-SAW-Delay} except~$\mathbb{T}=\SI{0.1\,}{\micro\second}$.}
	\label{F8-Qubits-Dynamics-via-SAW-Decay}
\end{figure}

\begin{figure}[b]
	\centering
	\includegraphics[scale=0.21]{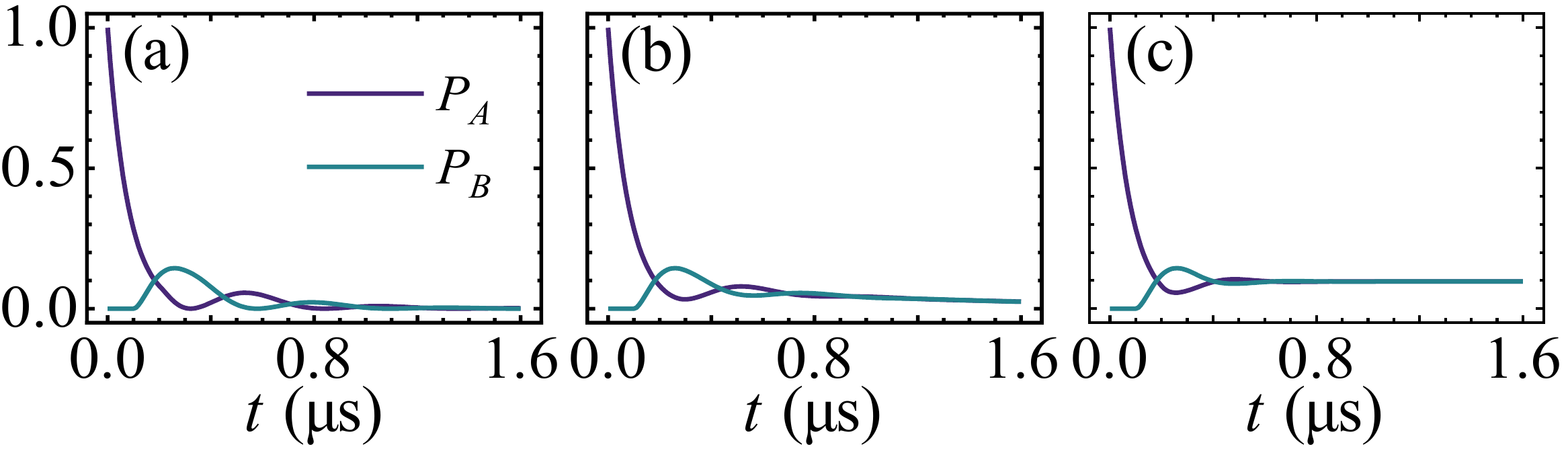}
	\caption{The population evolution of the two qubits when the relative phase between two qubits are taken as (a) $\theta_{\mathbb{T}}=\pi\big/2$, (b) $\theta_{\mathbb{T}}=3\pi\big/4$, and (c) $\theta_{\mathbb{T}}=\pi$. Other parameters are the same as those in Fig.~\ref{F7-Qubits-Dynamics-via-SAW-Delay} except $\mathbb{T}=0.1\,\SI{}{\micro\second}+{\rm mod}(\theta_{\mathbb{T}},\pi)/{\omega_{10}}$ and $\Gamma_0 /2\pi=1\,\text{MHz}$, with ${\rm mod}(\theta_{\mathbb{T}},\pi)$ denoting the remainder of $\theta_{\mathbb{T}}$ divided by $\pi$.}
	\label{F9-Qubits-Dynamics-via-SAW-Phase}
\end{figure}

In addition to the delay time, the decay rate $\Gamma_0$ of the qubit also plays a crucial role in the population evolutions of the two qubits. This is primarily because the decay rate $\Gamma_0$ determines the time for the energy transferred from the qubit to the SAW phonons in waveguide. We show the population evolutions of the two qubits for different decay rate $\Gamma_0$ in Fig.~\ref{F8-Qubits-Dynamics-via-SAW-Decay}. In the case of $\Gamma_0^{-1} \ll \mathbb{T}$ as shown in Fig.~\ref{F8-Qubits-Dynamics-via-SAW-Decay}(a), the decay time of the qubit is much smaller than the delay time. Thus, before the SAW phonons generated by the qubit A (qubit B) propagate to the qubit B (qubit A), the qubit A (qubit B) decays rapidly to the waveguide. As the decay rate decreases, e.g., $\Gamma_0^{-1} =\mathbb{T}$ as shown in Fig.~\ref{F8-Qubits-Dynamics-via-SAW-Decay}(b), the decay of the qubit becomes slower. Thus, before the qubit A (qubit B) decays completely to the waveguide, the qubit B (A) is excited by the SAW phonons that are generated by the qubit A (qubit B). Specially, in the case of $\Gamma_0^{-1} \gg \mathbb{T}$ as shown in Fig.~\ref{F8-Qubits-Dynamics-via-SAW-Decay}(c), the decay time of the qubit is much larger than the delay time. In such a case, before the qubit A (qubit B) decays completely to the waveguide, the phonons generated by the qubit A (qubit B) would be partly reflected back to the qubit A (qubit B) by the qubit B (qubit A). The reflected phonons interfere with the phonons generated by the decay of the qubit, which ultimately results in a portion of the energy being trapped within the two qubits.

From Fig.~\ref{F8-Qubits-Dynamics-via-SAW-Decay}(c), one can find that the decay of the two quibits to the waveguide is highly dependent on the interference between the phonons generated by the qubit A (qubit B) and the phonons reflected by the qubit B (qubit A). As shown in Eqs.~(\ref{A-Atom-dynamics})~and~(\ref{B-Atom-dynamics}), such an interference is determined by the phase $\theta_{T}=\omega_{10}\mathbb{T}$ with a period $\pi$. When the transition frequency $\omega_{10}$ of qubit is given, the interference is determined by the delay time $\mathbb{T}$. To clearly illustrate such an interference, we study the population evolutions of two qubits for different phase $\theta_{\mathbb{T}}$ for given frequency $\omega_{10}\big/2\pi=4.72\,\text{GHz}$ under the condition $\Gamma_0^{-1} \gg \mathbb{T}=\theta_{T}/\omega_{10}$, i.e., the decay time of the qubit is much larger than the delay time. When the qubit A (qubit B) decays into the waveguide, the qubit B (qubit A) can be considered as a mirror~\cite{Superconducting-Circuit4,Atom-Mirror-(anti)node1,Atom-Mirror-(anti)node2,Atom-Mirror-(anti)node3}, which leads to the standing wave for the SAW phonon with the frequency $\omega_k=\omega_{10}$. For the phase $\theta_{\mathbb{T}}=\pi/2$ as shown in Fig.~\ref{F9-Qubits-Dynamics-via-SAW-Phase}(a), each qubit is located at the antinode of the standing wave~\cite{Atom-Mirror-(anti)node1,Atom-Mirror-(anti)node2,Atom-Mirror-(anti)node3}, which results in the enhancement for the decay of qubit to the waveguide. As the phase $\theta_{\mathbb{T}}$ varies from $\pi/2$~to~$\pi$, each qubit is not located at the antinode of the standing wave, and thus the decay of qubit becomes slower, with e.g., $\theta_{\mathbb{T}}=3\pi/4$ as shown in Fig.~\ref{F9-Qubits-Dynamics-via-SAW-Phase}(b). Specially, for the phase $\theta_{\mathbb{T}}=\pi$ as shown in Fig.~\ref{F9-Qubits-Dynamics-via-SAW-Phase}(c), each qubit is located at the node of the standing wave~\cite{Atom-Mirror-(anti)node1,Atom-Mirror-(anti)node2,Atom-Mirror-(anti)node3}, which leads to the suppression for the decay of the qubit to the waveguide.

\begin{figure}[tp]
	\centering
	\includegraphics[scale=0.21]{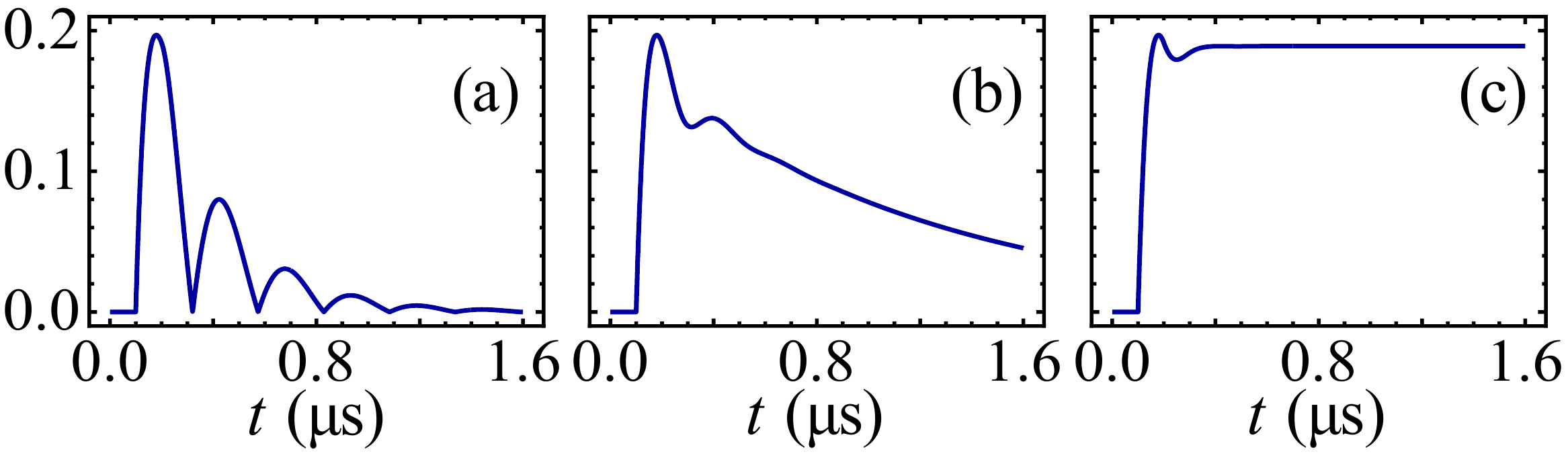}
	\caption{The concurrence of the two qubits when the relative phase between two qubits are taken (a) $\theta_{\mathbb{T}}=\pi\big/2$, (b) $\theta_{\mathbb{T}}=3\pi\big/4$, and (c) $\theta_{\mathbb{T}}=\pi$. Other parameters are the same as those in Fig.~\ref{F9-Qubits-Dynamics-via-SAW-Phase}.}
	\label{F10-Qubits-Entangle-via-SAW-Phase}
\end{figure}

Moreover, we point out that the phase $\theta_{\mathbb{T}}=\omega_{10}\mathbb{T}$ with period $\pi$ would also have effect on the entanglement of two qubits. In Fig.~\ref{F10-Qubits-Entangle-via-SAW-Phase}, we still focus on the condition $\Gamma_0^{-1} \gg \mathbb{T}$ and choose different phase $\theta_{\mathbb{T}}$ to show the evolution of the concurrence with the time $t$ for given frequency $\omega_{10}\big/2\pi=4.72\,\text{GHz}$. It is shown that the entanglement of two qubits reaches maximum $C\simeq0.2$ rapidly when both qubits are excited at time $t$ with $\mathbb{T} < t <2\mathbb{T}< \Gamma_{0}^{-1}$ with given frequency $\omega_{10}\big/2\pi=4.72\,\text{GHz}$. When the phase $\theta_{T}$ is taken as $\theta_{T}=\pi/2$ as considered in Fig.~\ref{F9-Qubits-Dynamics-via-SAW-Phase}(a), the enhanced decay of the qubit would lead to the rapid disentanglement of two qubits as shown in Fig.~\ref{F10-Qubits-Entangle-via-SAW-Phase}(a). When the phase $\theta_{\mathbb{T}}$ varies from $\pi/2$ to $\pi$, the enhancement for the decay of the qubit is modified, thus the decay from the entanglement to the disentanglement becomes slower as shown in Fig.~\ref{F10-Qubits-Entangle-via-SAW-Phase}(b) with, e.g., $\theta_{T}=3\pi/4$ as considered in Fig.~\ref{F9-Qubits-Dynamics-via-SAW-Phase}(b). In particular, when the phase $\theta_{\mathbb{T}}$ is taken as $\theta_{\mathbb{T}}=\pi$ as considered in Fig.~\ref{F9-Qubits-Dynamics-via-SAW-Phase}(c), the suppressed decay of qubit would ultimately result in the steady entanglement with the concurrence $C \simeq 0.19$ as shown in Fig.~\ref{F10-Qubits-Entangle-via-SAW-Phase}(c).

\section{Conclusion}
\label{Sec6-Conclusion}

In conclusion, we have studied the quantum theory for SAW in the piezomagnetic medium by resorting to the canonical quantization method. Based on the theory of the classical SAW, which is obtained by solving the general wave equation in the piezomagnetic medium, we derive the Hamiltonian of the quantized SAW in the strip waveguide made of piezomagnetic medium, and give the quantization expressions for the mechanical displacement and magnetic field induced by piezomagnetic effect. Based on this, we study the interaction between several typical magnetic quantum systems (i.e., supercondicting qubits, ferromagnnetic magnons, and defect centers in diamond) and the quantized single-mode SAW in piezomagnetic waveguide at single-phonon level. It is found that the intrinsic properties of SAW in piezomagnetic medium enable it interact with these magnetic systems via the magnetic field or strain induced by piezomagnetic effect. This is very different from the SAW in piezoelectric medium, which is coupled to quantum systems via the electric field or strain induced by piezoelectric effect. Furthermore, we study the interaction between two qubits mediated by multi-mode SAW phonon in piezomagnetic waveguide. The results show that the interaction between two qubits depends strongly on the delay time, decay of the qubit, and phase due to the propagation of SAW between two qubits via the waveguide. This leads to the time-delay Rabi oscillation, energy trapping within the qubits, and steady entanglement between two qubits. We mention that the coupling strength between the magnetic quantum systems and quantized single-mode SAW in the piezomagnetic waveguide is not strong at single-phonon level. To achieve the strong-coupling, a SAW resonator fabricated on the surface of the piezomagnetic medium needs to be further explored.

In our present models, since the magnetic quantum systems (used as qubits) are coupled to the SAW phonons via the magnetic field or stain induced by piezomagnetic effect, the information of SAW phonons can be detected by the qubits. Thus, when the external disturbances (e.g., magnetic field, strain, or temperature) is applied to the piezomagnetic medium, one could realize the qubit-mediated sensors with SAW phonons in piezomagnetic medium. As a prospect, by designing the interaction between the qubits and SAW phonons, one can prepare the nonclassical states (e.g., squeezed state or entangled state) of SAW phonons to achieve the precision measurement. Therefore, our studies would provide new ways for the quantized SAW-based sensing. Furthermore, besides the magnetic quantum systems discussed in our present studies, the quantized SAW in piezomagnetic medium can also be coupled with other quantum systems, such as quantum dots~\cite{SAW-QuantumDot-Coupling1,SAW-QuantumDot-Coupling2,SAW-QuantumDot-Coupling3}, chiral molecules~\cite{ChiralMolecules-SAW-Strain,ChiralMolecules-SAW-Piezoelectric}, and waveguides~\cite{SAW-Waveguide1,Waveguide-SAW-Review,Waveguide-SAW-Routing}. Therefore, the quantized SAW in piezomagnetic medium provides an alternative way for the integration of different quantum systems and may have potential applications in the quantum communication and quantum computing.

\begin{acknowledgments}
This work was supported by the National Natural Science Foundation of China with Grants No.~12374483 and No.~92365209.
\end{acknowledgments}

\end{document}